\documentclass[twocolumn]{article}

\usepackage{arxiv}
\usepackage{multicol}
\usepackage{amssymb}
\usepackage{amsmath}
\usepackage{subcaption}
\usepackage{booktabs}   
\usepackage{tabularx}  
\usepackage{float}
\usepackage{orcidlink}
\usepackage{graphicx}
\usepackage{dblfloatfix}
\usepackage[numbers,sort&compress]{natbib}
\usepackage[T1]{fontenc}
\usepackage{makecell}

\title{QTIS: A QAOA-Based Quantum Time Interval Scheduler}

\author{ \orcidlink{0009-0008-1726-6506} {Jos{\'e} A. Tirado-Dom{\'i}nguez} \\
    Department of Computer Architecture \\
    Universidad de M\'{a}laga \\
    M\'{a}laga, Spain \\
	\texttt{jatirado@uma.es} \\
	\And
	\orcidlink{0000-0001-9748-9161}{Eladio Guti{\'e}rrez} \\
	Department of Computer Architecture \\
    Universidad de M\'{a}laga \\
    M\'{a}laga, Spain \\
	\texttt{eladio@uma.es} \\
    \And
	\orcidlink{0000-0003-2233-0011}{Oscar Plata} \\
	Department of Computer Architecture \\
    Universidad de M\'{a}laga \\
    M\'{a}laga, Spain \\
	\texttt{oplata@uma.es}
}

\renewcommand{\headeright}{}
\renewcommand{\undertitle}{}
\renewcommand{\shorttitle}{Quantum Time Interval Scheduler}

\hypersetup{
pdftitle={QTIS: A QAOA-Based Quantum Time Interval Scheduler},
pdfsubject={Quantum Physics (quant-ph)},
pdfauthor={Jose A. Tirado-Dominguez, Eladio Gutierrez, Oscar Plata},
pdfkeywords={Quantum Approximate Optimization Algorithm (QAOA),
Variational Quantum Algorithms (VQA),
Hybrid Quantum–Classical Computing,
Task Scheduling,
Quadratic Unconstrained Binary Optimization (QUBO),
Quantum Circuit Design,
Constraint Enforcement,
Quantum Annealing,
Time-Interval Optimization}
}

\begin{document}

\twocolumn[
\begin{@twocolumnfalse}

\maketitle

\begin{abstract}
Task scheduling with constrained time intervals and limited resources remains a fundamental challenge across domains such as manufacturing, logistics, cloud computing, and healthcare. 
This study presents a novel variant of the Quantum Approximate Optimization Algorithm (QAOA) designed to address the task scheduling problem formulated as a Quadratic Unconstrained Binary Optimization (QUBO) model.
The proposed method, referred to as Quantum Time Interval Scheduler (QTIS), integrates an ancilla-assisted quantum circuit to dynamically detect and penalize overlapping tasks, enhancing the enforcement of scheduling constraints. Two complementary implementations are explored for overlap detection: a quantum approach based on RY rotations and CCNOT gates, and a classical alternative relying on preprocessed interval comparisons. 

QTIS decomposes the problem Hamiltonian, $H_P$, into two components, each parameterized by a distinct angle. The first component encodes the objective function, while the second captures penalty terms associated with overlapping intervals, which are controlled by the auxiliary circuit.

Subsequently, three minimization strategies are evaluated: standard QAOA, T-QAOA, and HT-QAOA, showing that employing separate parameters for the different components of the problem Hamiltonian leads to lower energy values and improved solution quality. 
Results confirm the efficiency of QTIS in scheduling tasks with fixed temporal windows while minimizing conflicts, demonstrating its potential to advance hybrid quantum–classical optimization in complex scheduling environments.

\end{abstract}
\vspace{0.5em}
\vspace{2em}
\end{@twocolumnfalse}
]

\section{Introduction}\label{sec:Introduction}

Task scheduling is a fundamental optimization problem with applications in multiple real-world domains such as industry~\cite{HosseiniAmir2024Simw,MomenikorbekandiAtefeh2025ISMf}, logistics and transportation~\cite{LiuXiaobo2023ASLR}, computation~\cite{González-San-MartínJessica2024ARoS,AvanAmin2023ASRo}, or healthcare~\cite{AbdalkareemZahraaA.2021Hsio,LiuZhuo2025RoHH}, among others. 

Task scheduling has traditionally been modeled with the classical Travelling Salesman Problem (TSP)~\cite{Dantzig1954,Lawler1985}  or the Job Shop Problem (JSP)~\cite{Holt1955} with the objective of determining the optimal sequence and assignment of tasks to resources to satisfy predetermined objectives, such as minimizing cost, maximizing throughput, or ensuring timely completion~\cite{alma991010696818904986}.

These problems belong to the class of NP-hard combinatorial optimization problems~\cite{GareyM.R1976TCoF} and, as noted by Graham et al.~\cite{Graham1979}, finding a guaranteed optimal solution becomes computationally intractable as the problem size increases. This inherent complexity necessitates the use of approximation methods, such as heuristics~\cite{Lin1973,Adams1988} and meta-heuristics~\cite{Gutin2002,Nowicki1996,zhang2025metaheuristics,van2021discovering}, for solving real-world, large-scale instances in a reasonable time frame.

Although classical algorithms can accelerate the search for solutions, optimization problems may also benefit from advanced computing paradigms, particularly quantum computing, which offers promising approaches for managing complex scenarios in high-concurrency environments ~\cite{MontanaroAshley2016Qaao, FarhiEdward2014AQAO} .

 In this sense, Adiabatic Quantum Computing (AQC)~\cite{AlbashTameem2018Aqc, Crosson2021_NatRevPhys} is a paradigm particularly suited to tackling combinatorial optimization problems~\cite{CarugnoCostantino2022Etjs, Ossorio-CastilloJ.2022Ooar}.
 In AQC, the optimization problem is encoded into a Hamiltonian whose ground state corresponds to the optimal solution. The system is initially prepared in the ground state of a simple Hamiltonian and then evolved slowly enough—according to the adiabatic theorem—to the ground state of the problem Hamiltonian.~\cite{BornFock1928}.

In addition to adiabatic quantum computing, other paradigms of quantum computation, such as gate-based quantum computing, have proven effective in addressing combinatorial optimization problems~\cite{ChicanoFrancisco2025Cowq, AmaroDavid2022Acso,SLYSZ2026107934} . These universal quantum computing systems operate by executing quantum circuits composed of gates that implement unitary operators on a set of qubits. The coherent and controlled manipulation of qubits enables the implementation of quantum algorithms capable of surpassing classical methods in certain optimization contexts~\cite{PeruzzoAlberto2014Aves,FarhiEdward2014AQAO}. 

Despite the theoretical promise of quantum advantage, the practical deployment of quantum algorithms for real-world applications is currently constrained by the limitations of available hardware in the Noise-Intermediate-Scale Quantum (NISQ) devices~\cite{PreskillJohn2018QCit, BhartiKishor2022Niqa}.
These systems are highly sensitive to noise, still suffer from short decoherence times, and are limited in the number of qubits and the depth of circuits they can reliably execute. Consequently, their inherent limitations prevent them from independently solving large-scale optimization problems, thereby motivating the development of hybrid classical–quantum approaches~\cite{GE2022314,CallisonAdam2022Hqai}.

These hybrid approaches, that combine quantum algorithms with classical computational resources, exploit the strengths of both paradigms and have demonstrated clear advantages over purely classical approaches. In particular, variational quantum algorithms (VQAs) have become leading candidates for optimization tasks in the NISQ era~\cite{Cerezo2021}.
These hybrid methods rely on parameterized quantum circuits to generate candidate solutions, while a classical optimizer iteratively updates the parameters to minimize a predefined cost function. Through this feedback loop, the quantum–classical system progressively steers the circuit toward an approximation of the problem’s optimal solution.

Two of the most widely studied variational algorithms are the Variational Quantum Eigensolver (VQE)~\cite{PeruzzoAlberto2014Aves} and the Quantum Approximate Optimization Algorithm (QAOA)~\cite{FarhiEdward2014AQAO, BlekosKostas2024AroQ}. 

The VQE algorithm is designed to determine the lowest eigenvalue and corresponding eigenstate of a system’s Hamiltonian. Originally developed for quantum chemistry and materials science~\cite{McArdleSam2020Qcc,KandalaAbhinav2017Hvqe}--where a molecule’s or material’s ground state governs its properties--, VQE has since been extended to continuous optimization, graph-theoretic problems, and various industrial applications. The method relies on the variational principle of quantum mechanics, which guarantees that the expectation value of a Hamiltonian for any trial state (ansatz)--\(C=\left\langle\Psi(s)|H|\Psi(s)\right\rangle\)-- provides an upper bound to the true ground-state energy.

Quantum Approximate Optimization Algorithm (QAOA) \cite{FarhiEdward2014AQAO} is a leading variational algorithm designed for combinatorial optimization problems, such as MaxCut, TSP, VRP, and Job-Shop Scheduling. It approximates the optimal solution by emulating adiabatic evolution through a discrete sequence of unitary operators constructed from a problem Hamiltonian $H_P$ and a mixer Hamiltonian $H_B$, parameterized by angles $\gamma$ and $\beta$.  A classical optimizer iteratively updates these parameters to minimize the expected energy, guiding the system toward the ground-state configuration of $H_P$, which encodes the desired solution.
Various QAOA variants have been proposed to enhance performance, scalability, and applicability, including Multi-angle QAOA~\cite{herrman2021multiangle}, XQAOA~\cite{vijendran2023xqoaa}, QAOA+~\cite{wang2023quantum}, Two-Step QAOA~\cite{minato2024two}, and the Quantum Alternating Operator Ansatz~\cite{hadfield2019qaoa}.

In this work, we propose a novel version of the QAOA algorithm designed to address a particular case of task scheduling characterized by strict time restrictions for both start times and durations. 

Formally, the problem is defined as follows:

\begin{figure}[!t]
    \centering
    \includegraphics[width=0.9\columnwidth]{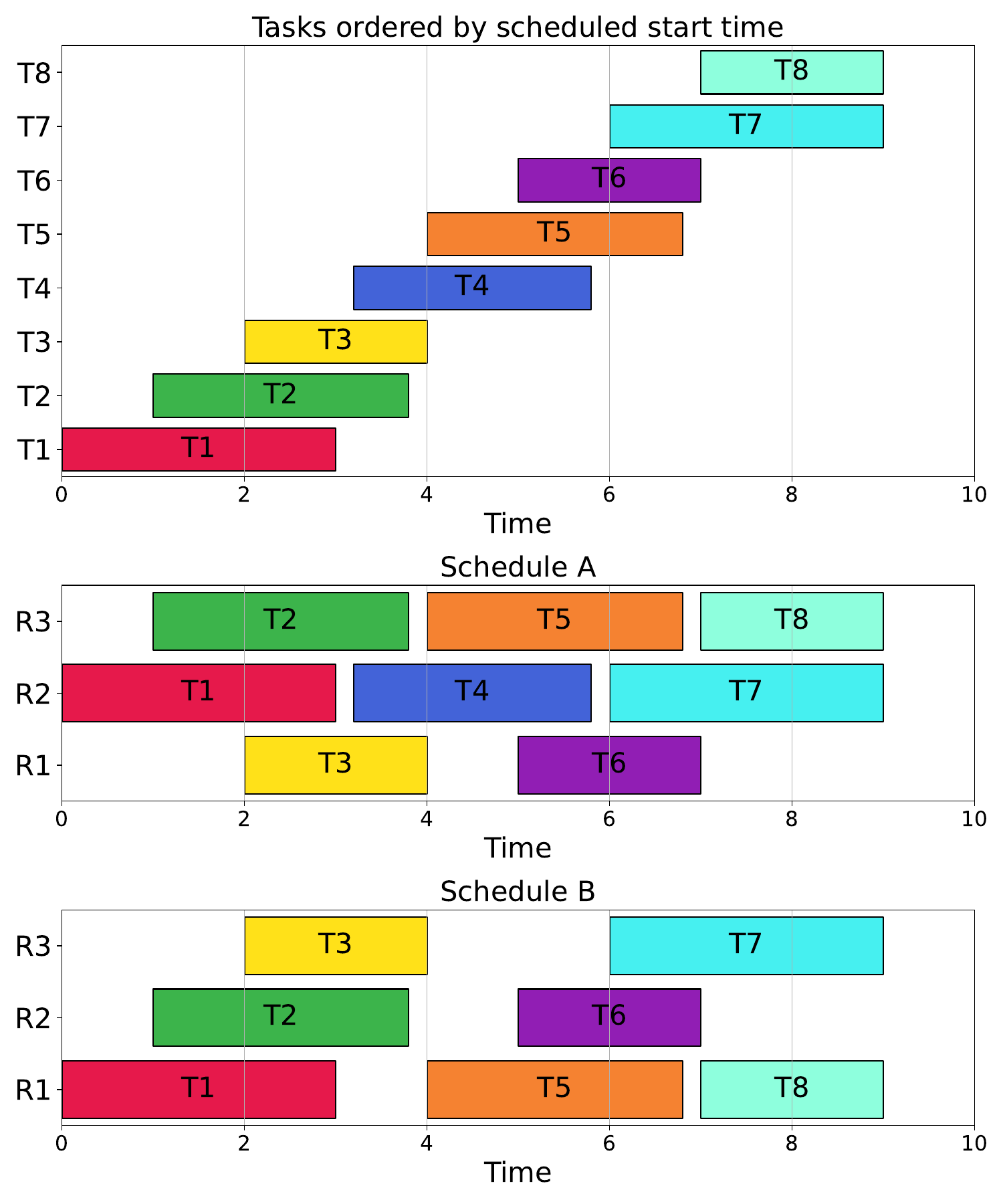}
    \caption{Task scheduling problem with start time and duration restrictions. Schedule A with optimal planning for the eight tasks, Schedule B with no allocated T4 task}
    \label{fig:qtisproblem}
\end{figure}

\textit{Given a set of tasks with predefined durations and start times, along with a limited set of resources capable of running only one task at a time, the objective is to allocate the maximum number of tasks while avoiding conflicts.} 

An example of the problem is illustrated in Fig.~\ref{fig:qtisproblem}. A set of eight tasks \((T_1\dots T_9)\), each with different start times and durations, must be scheduled using three resources 
\((R_1,R_2,R_3)\). The feasibility of the schedule depends on the resource assignment: in Schedule A, all tasks are successfully executed, whereas in Schedule B, some tasks cannot be completed.

This scheduling problem, involving fixed time intervals and limited resources, appears in many practical contexts. Examples include industrial maintenance, just-in-time production, aircraft take-off and landing, public transport, computing clusters, cloud job allocation, and surgical procedure scheduling. In all cases, tasks must be assigned to resources within predefined time windows while respecting capacity constraints. Efficient scheduling is essential to optimize resource utilization and minimize delays or downtime. 
The diverse range of these applications demonstrates the prevalence of time-constrained scheduling problems and motivates the exploration of quantum computing approaches to achieve more effective optimization.

To address this scheduling problem, we formulate it as a Quadratic Unconstrained Binary Optimization (QUBO) problem, encoding time and resource constraints into a cost function whose feasible solutions represent approximately optimal task allocations.

Unlike traditional QAOA, our approach introduces an ancilla-assisted quantum circuit that explicitly handles task collisions and transfers them into the ansatz. In this case, the ansatz involves dividing the problem's Hamiltonian into two different Hamiltonians.
The first is dedicated to coding the problem and assigning it to resources, and the second is in charge of handling the constraint that two tasks cannot be executed simultaneously on the same resource.

\noindent The main contributions of this work are summarized bellow:

First, the task scheduling problem with strict temporal constraints is formulated as a Quadratic Unconstrained Binary Optimization (QUBO) model, allowing its representation within a quantum optimization framework. 
Second, a quantum circuit for time-collision detection is defined, incorporating two alternative schemes tailored to different scheduling contexts. 
Third, a novel variant of the standard Quantum Approximate Optimization Algorithm (QAOA), referred to as Quantum Time Interval Scheduler QAOA (QTIS-QAOA), is proposed. This formulation introduces three Hamiltonians, \(H_p\), \(H_c\), and \(H_B\), where \(H_p\) encodes the optimization objective, \(H_c\) enforces the time-collision constraints, and \(H_B\) represents the standard mixer Hamiltonian. 
Furthermore, it is demonstrated that the ansatz circuit can be constructed using elements derived from an ancillary quantum subcircuit, effectively embedding constraint information within the variational form. 
It is also shown that employing a distinct parameter set for the \(H_c\) Hamiltonian, rather than sharing the same \(\gamma_i\) parameters applied to \(H_p\), enhances the optimization performance of the QTIS-QAOA algorithm. 
Finally, a parameter initialization method for the ansatz circuit,HT-QAOA, is introduced, enabling higher solution accuracy without significantly increasing computational cost.

The remainder of this paper is organized as follows:
Section~\ref{sec:Background} provides a background on QAOA and related work. Section~\ref{sec:Formulation} presents the problem formulation and its transformation into a QUBO model. Section~\ref{sec:QTIS} describes the principles of the modified QAOA algorithm. Section~\ref{sec:Conflict} illustrates the role of the auxiliary quantum circuit in enforcing constraints. Section~\ref{sec:Simulations} discusses experimental results and performance analysis. Finally, Section~\ref{sec:Conclusion} concludes with key insights and directions for future research.

\section{Background and related work}
\label{sec:Background}

\subsection{QAOA}

The Quantum Approximate Optimization Algorithm (QAOA) is an optimization algorithm inspired by the adiabatic theorem \cite{FarhiEdward2000QCbA}.
\begin{figure*}[ht]
    \centering    \includegraphics[width=\textwidth]{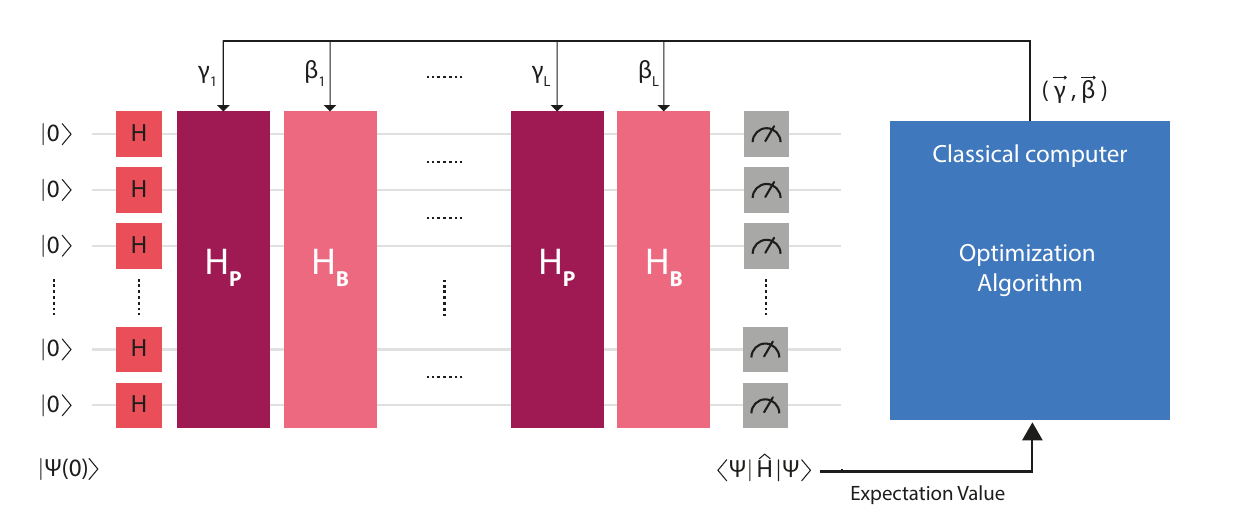}
    \caption{QAOA Algorithm}
    \label{fig:classicqaoa}
\end{figure*}
It is grounded in the time-dependent Schrödinger equation, which governs the temporal evolution of a quantum system's state under a Hamiltonian such that:

\begin{equation}
i\hbar \frac{\partial}{\partial t} \left| \Psi(t) \right\rangle = \hat{H}(t) \left| \Psi(t) \right\rangle
\label{eq:schrodinger}
\end{equation}

If \(H(t)\) is time independent, the solution of equation~\ref{eq:schrodinger} takes the form:

\begin{equation}
    \left|\Psi(t)\right\rangle = \sum_n c_n(0)e^{i\frac{E_n(0)}{\hbar}t}\left|n(0)\right\rangle
\end{equation}
where \(E_n(0)\) and \(\left|n(0)\right\rangle\) are the eigenvalues and eigenvectors of the initial Hamiltonian defined by Eq.~\ref{eq:eigenvalues}.

\begin{equation}
    \hat{H}(0)\left|\Psi(0)\right\rangle = En(0)\left|n(0)\right\rangle\
\label{eq:eigenvalues}    
\end{equation}

The adiabatic theorem states that, when the evolution of \(H(t)\) is sufficiently slow, if the system is initially in the ground state \(\left|\Psi(0)\right\rangle\), then at time \(T\), for large enough \(T\), the system will stay in the new ground state \(\left|\Psi(T)\right\rangle\), characterized by the eigenvalues and eigenvectors of the final Hamiltonian \(\hat{H}(T)\):

\begin{equation}
    \hat{H}(T)\left|\Psi(T)\right\rangle= E_n(T)\left|n(T)\right\rangle
\end{equation}

This statement allows for the construction of an interpolating Hamiltonian of the form \(H(s) = (1-s)H_B + sH_P\), with \(s = \frac{t}{T}\), where \(H_B\) is an initial base Hamiltonian and \(H_P\) is a problem-dependent Hamiltonian. The adiabatic evolution of the system drives it toward a final ground state encoding the solution to the optimization problem.

QAOA leverages the adiabatic theorem to approximate the solution of an optimization problem by defining two unitary operators based on the Hamiltonians \(H_P\) and \(H_B\), parameterized by the angles \(\gamma\) and \(\beta\):

\begin{equation}
    U (H_P, \gamma) = e^{-i\gamma H_P}
\end{equation}
\begin{equation}
    U (H_B, \beta) = e^{-i\beta H_B}
    \label{eq:uhb}
\end{equation}

The Hamiltonian of the problem \(H_P\) is constructed from the optimization objective function, which is typically formulated as a Quadratic Unconstrained Binary Optimization (QUBO) problem and then mapped to the Ising model. In contrast, \(H_B\) is a fixed mixing Hamiltonian, expressed as:

\begin{equation}
    H_B = \sum_{j=1}^n \sigma^x_j
    \label{eq:HB}
\end{equation}

The QAOA algorithm builds the ansatz by alternating applications of operators \(U(H_P, \gamma_i)\) and \(U(H_B, \beta_i)\), repeated \(L\) times, as shown in Fig.~\ref{fig:classicqaoa}.

Before applying the operators, the initial quantum state \(\left|\Psi(0)\right\rangle = \left|+\right\rangle\) is prepared by applying a Hadamard gate to each qubit in the register.

Once the quantum circuit is constructed, a classical optimization algorithm is employed to iteratively adjust the set of parameters \(\{\gamma_i, \beta_i\}\) in order to minimize the expectation value:

\begin{equation}
\left\langle \gamma, \beta \right| H_P \left| \gamma, \beta \right\rangle
\end{equation}

where:

\begin{equation}
\begin{split}
    \left| \gamma, \beta \right\rangle = & U(H_B, \beta_L)U(H_P, \gamma_L)\dots \\
    &U(H_B, \beta_L) U(H_P, \gamma_1) \left|s\right\rangle 
\end{split}
\end{equation}

and the initial state \(\left|s\right\rangle\) is defined as:

\begin{equation}
    \left|s\right\rangle = \left|+_1\right\rangle \dots \left|+_n\right\rangle
\end{equation}

\subsubsection{\( \vec{\gamma}\)  and  \(\vec{\beta}\) initialization}

The proper initialization of the parameters $\vec{\gamma}$ and $\vec{\beta}$ is crucial for the rapid convergence of the classical optimization algorithm. An appropriate choice of these parameters can significantly reduce the associated computational complexity.

Numerous studies~\cite{LeeXinwei2023ADIS,AmosyOhad2024Iqao,ZhouLeo2020QAOA,KunduAkash2024Hhqa} have proposed different strategies for the initial parameter selection. The simplest approach consists of choosing $\vec{\gamma}$ randomly within the range $[0, 2\pi]$, and $\vec{\beta}$ within $[0, \pi]$. More sophisticated methods rely on heuristic interpolation techniques to estimate the initial parameters for a circuit of greater depth based on the optimization results from a shallower circuit with depth $L = 1$ or $2$~\cite{LeeXinwei2023ADIS}. For specific cases such as MaxCut, strategies based on neural networks have been proposed~\cite{AmosyOhad2024Iqao}, which can approximate the final parameters with high accuracy, effectively eliminating the need for a classical optimization loop.

\subsection{Related work}

The Quantum Approximate Optimization Algorithm (QAOA) was originally proposed to solve the MaxCut problem, an NP-hard combinatorial optimization problem. Given a graph \( G = (V, E) \), the goal of MaxCut is to partition the set of vertices \( V \) into two disjoint subsets \( S \) and \( \overline{S} \), such that \( V = S \cup \overline{S} \), in a way that maximizes the number of edges between elements in \( S \) and those in \( \overline{S} \).

Beyond MaxCut, QAOA has been explored as a heuristic for other optimization problems classified as NP-Hard, such as the Travelling Salesman Problem (TSP)~\cite{Qian2023} or Graph partitioning~\cite{Chukwu2020} and colouring~\cite{Bravyi2022} among others.  

In the context of task scheduling, the application of the Quantum Approximate Optimization Algorithm to solve the Job Shop Scheduling Problem (JSSP) has proven to be highly effective in domains such as logistics~\cite{DalalArchismita2024Dcqa} or manufacturing scenarios~\cite{AmaroDavid2022Acso}.

\begin{figure}[ht]
    \centering
    \includegraphics[width=0.9\columnwidth]{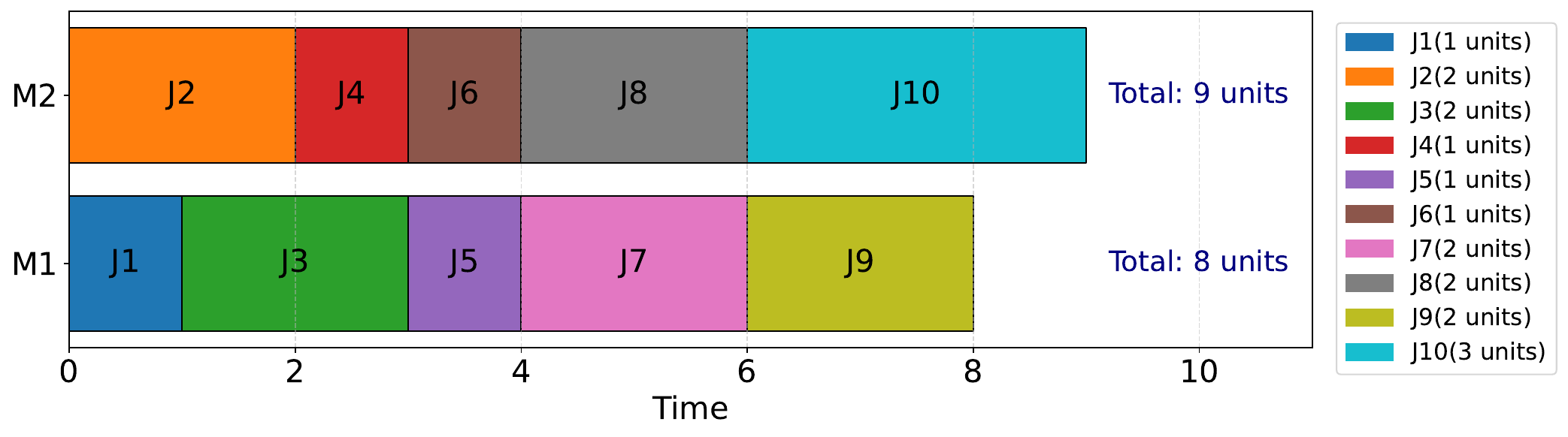}
    \caption{Example of Job Shop Scheduling Problem with 10 jobs and 2 Machines. The goal of the problem is minimizing the total processing time, in this case 9 time units.}
    \label{fig:jssp}
\end{figure}

The Job Shop Scheduling Problem (illustrated in Fig.~\ref{fig:jssp}) involves assigning a set of jobs \( J = \{J_1, J_2, \dots, J_n\} \) with different processing times to a set of machines \( M = \{M_1, \dots, M_k\} \) with varying processing capacities, with the objective of minimizing the total processing time. Some variants of the JSSP include job dependencies or machines specialized in certain types of tasks ~\cite{DalalArchismita2024Dcqa}.

In other areas, the task scheduling problem needs to take into account temporal constraints such as deadlines or time execution windows. In this context, Khalid et al.~\cite{Khalid2024} propose the use of QAOA to address a task scheduling problem for 6G networks in the cloud radio access network. In their formulation, tasks are processed at the Remote Radio Heads, each with an associated execution deadline. Temporal constraints are also particularly critical in deep space applications. As noted by Geda et al.~\cite{geda2025adaptive}, QAOA can be applied to schedule satellite imaging tasks, ensuring that areas of the Earth are imaged within the bounded time windows when the satellite is in orbit over them.

\section{Problem formulation}
\label{sec:Formulation}

Our problem involves scheduling the execution of a series of tasks with a limited set of resources, considering the following rules:  

\begin{enumerate}
    \item  Each task has a fixed execution duration and a predefined time window, determined by a start time and an end time $(t_{end} = t_{start} + duration)$, which cannot be arbitrarily modified.  
    \item The available execution resources are finite, and each resource can execute only one task at any given time.
    \item Some tasks may overlap in time, preventing them from being executed on the same resource simultaneously.
\end{enumerate}

With these premises, the goal is to maximize the number of tasks that can be executed with the available set of resources. 

\subsection{QUBO}
To translate this problem into a QUBO formulation, we will define a set of binary variables $x_{ij}$ where: 

\begin{equation}
x_{ij} = \begin{cases}
1 & \text{if task i is scheduled in resource j} \\
0 & \text{otherwise}
\end{cases}
\end{equation}

With this definition, our goal is equivalent to minimizing the function \eqref{fobjetive} with a set of constraints. 

\begin{equation}
f(x) = -\sum_{i}\sum_{j}x_{ij}
\label{fobjetive}
\end{equation}

The first constraint corresponds to the fact that tasks can only be scheduled once in one of the available resources. 

\begin{equation}
    \sum_{j}x_{ij} = 1 \quad \forall i
    \label{fconstraint1}   
\end{equation}

The second constraint avoids scheduling two overlapped tasks on the same resource. That is,

\begin{equation}
    x_{ij}x_{kj}c_{ik} = 0 \quad \forall i,j,k>i
    \label{fconstraint2}
\end{equation}

where the overlap coefficient $c_{ik}$ can be defined as :

\begin{equation}
c_{ik} = \begin{cases}
1 & \text{if task i and k are overlapped} \\
0 & \text{otherwise}
\end{cases}
\end{equation}

With all these definitions, we can build the QUBO to minimize as  

\begin{equation}
\begin{split}
    \mathit{QUBO} = & -\sum_{i}\sum_{j}x_{ij} \\
    & + \sum_iP(\sum_{j}x_{ij} - 1)^2 \\
    & + \sum_j\sum_k\sum_iP(x_{ij}*x_{kj}*c_{ik})^2
\end{split}
\label{eq:fnqubo}
\end{equation}

The penalty factor \textit{P} has to be a number larger than the maximum value of $|f(x)|$ 

\begin{equation}
    P = I \cdot J + 1
\end{equation}

where $I = \textit{Number of tasks}$ and $J = \textit{Number of resources}$.

\section{QTIS-QAOA}
\label{sec:QTIS}

\begin{figure*}[ht]
    \centering
    \includegraphics[width=\textwidth]{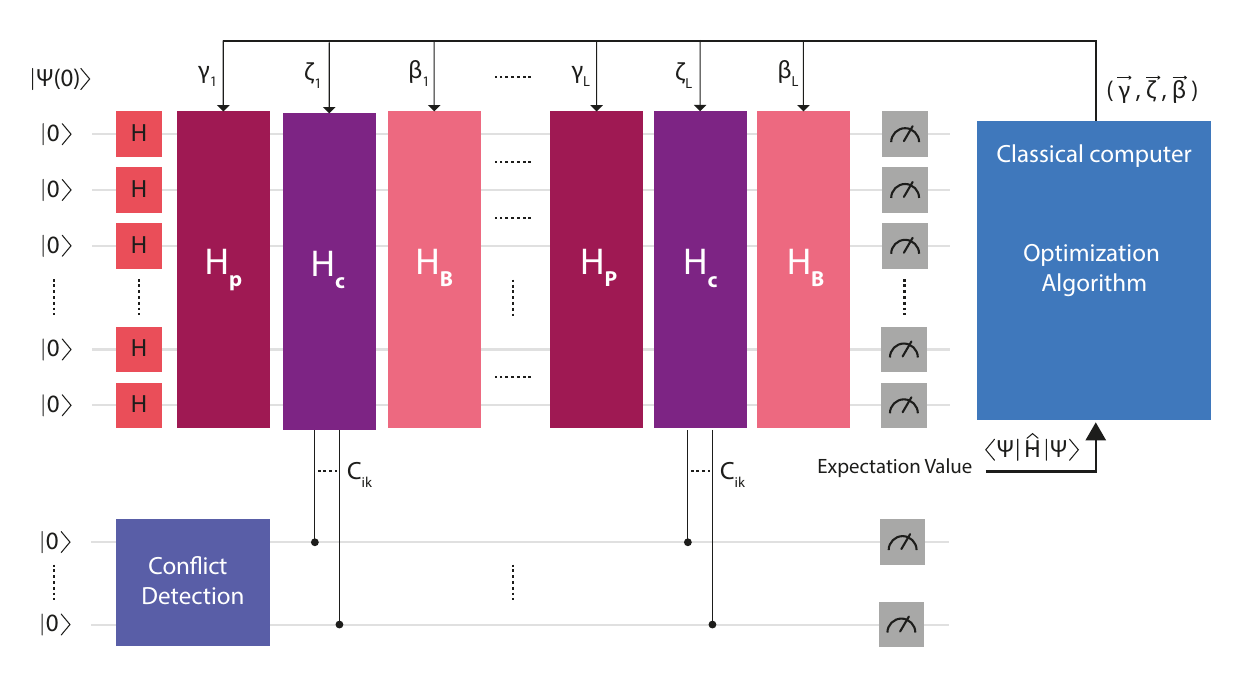}
    \caption{L layers QTIS-QAOA with the parametrized circuits associated to the Hamiltonians \(H_p, H_c, H_B\) and the conflict detection circuit}. 
    \label{fig:qtis-qaoa}
\end{figure*}

In continuation of the formulation presented in Section~\ref{sec:Formulation}, this section introduces QTIS (Quantum Time Interval Scheduler), a modified version of the standard QAOA algorithm designed to address the corresponding task scheduling problem.

As outlined in the theoretical background, QAOA operates by alternating two parameterized unitary operators, \(U(H_P, \gamma_i)\) and \(U(H_B, \beta_i)\), derived from the problem Hamiltonian \(H_P\) and the mixer Hamiltonian \(H_B\), respectively, over a circuit of depth \(L\). 
At each layer, a distinct pair of parameters \((\gamma_i, \beta_i)\) is optimized through a classical routine to minimize the expected value of the cost Hamiltonian.

The problem Hamiltonian \( H_P \) can be obtained by transforming the QUBO formulation of the scheduling problem into its equivalent Ising representation, while \( H_B \) remains fixed and corresponds to the definition provided in equation~\eqref{eq:HB}. 

The construction of the \(U(H_P, \gamma_i)\) ansatz requires complete pre-computation of the corresponding Ising Hamiltonian, which prevents the inclusion of elements that could otherwise be computed dynamically within the quantum circuit.

To overcome this limitation, the proposed QTIS-QAOA introduces an extended formulation in which \(H_P\) is partitioned into two components, \(H_p\) and \(H_c\), as follows:

\begin{equation}
    H_P = H_p + H_c
\end{equation}

In this formulation, \( H_p \) is derived from the objective function together with a subset of the penalty terms, while \(H_c\) encapsulates the remaining penalties. The application of these latter terms is controlled by ancillary qubits generated through a complementary quantum circuit, as discussed in the following sections.

Under this decomposition, the operator \(U(H_P, \gamma_i)\) is correspondingly divided into two operators \(U(H_p, \gamma_i)\) and \(U(H_c, \zeta_i)\). Each parameterized with different angles \(\gamma\) and \(\zeta\), respectively. The resulting QTIS-QAOA ansatz is then composed of a sequence of L layers alternating the three operators  \(U(H_p, \gamma_i)\), \(U(H_c, \zeta_i)\), and \(U(H_B, \beta_i)\) as illustrated in Fig.~\ref{fig:qtis-qaoa}. 

\subsection{\( H_p \)}

\( H_p \) and \(H_c\) can be obtained by dividing the QUBO function ~\eqref{eq:fnqubo} representing our problem into two parts. The first part, corresponding to the \( H_p \) Hamiltonian,  includes the objective function~\eqref{fobjetive} and the first set of penalties~\eqref{fconstraint1} related to the constraint that tasks can
only be scheduled once in one of the available resources. That is, 
\begin{equation}
   \mathit{QUBO_{H_p}} = -\sum_{i}\sum_{j}x_{ij}  + \sum_iP(\sum_{j}x_{ij} - 1)^2
\end{equation}

The transformation of \(\mathit{QUBO_{H_p}}\) into the Ising model gives us the final expression of \( H_p \), which takes the form:

\begin{equation}
H_p = \sum_{n}\sum_{m!=i}J_{nm}s_{n}s_{m} + \sum_{n}J_{nn}s_{n}
\label{fhpising}
\end{equation}

The \( H_p \) Hamiltonian can then be used to define the operator \(U(H_p, \gamma_i)\) as: 

\begin{equation}
    U (H_p, \gamma_i) = e^{-i\gamma_i H_p}
\end{equation}    

This formulation of \(U(H_p, \gamma_i)\) allows us to realize the operator as an ansatz circuit, composed of a set of quantum gates \(Rzz\) and \(Rz\) which are parametrized by $\gamma_i$ and the coefficients \(J_{nm}, J_{nn}\).

\begin{equation}
R_{Z_{n}Z_{m}}(2\gamma_lJ_{nm})^{\otimes ^{nm}}\otimes R_{Z_n}(2\gamma_lJ_{nn})^{\otimes ^n}
\end{equation}

\subsection{\(H_c\)}
 
The Hamiltonian \(H_c\) is derived from the second set of constraints~\eqref{fconstraint2}, which correspond to the remaining penalty terms in the original QUBO formulation~\eqref{eq:fnqubo}.
These constraints are designed to prevent tasks from conflicting by occupying the same resource simultaneously.

\begin{equation}
    \mathit{QUBO_{H_c}} =  \sum_j\sum_k\sum_iP(x_{ij}*x_{kj}*c_{ik})^2
\label{eq:fnqubohc}
\end{equation}

The \( H_c \) Hamiltonian can be computed, as shown in Appendix A, by applying the Ising transformation\footnote{Ising transformation:  $x_{ij} = \frac{(1-s_{ij})}{2}$ , $x_{kj} = \frac{(1-s_{kj})}{2}$}  to the previous \(\mathit{QUBO_{H_c}}\) equation.  
Consequently, the final form of \( H_c \) is given by:

\begin{equation}
    H_c = \sum_j\sum_k\sum_i\frac{P}{4}(1-s_{ij}-s_{kj}+s_{ij}s_{kj})c_{ik}
\end{equation}

Similarly to how we proceeded with \(H_p\), the operator \(U(H_c, \zeta_i)\) can be obtained as: 

\begin{equation}
    U (H_c, \zeta_i) = e^{-i\zeta_i H_c}
\end{equation}

The ansatz circuit for \(U(H_c, \zeta_i)\) can be constructed with a set of \(R_{zz}\) and \(R_{z}\) gates,  following the same method used for \(U(H_p, \gamma_i)\), as follows:

\begin{equation}
\begin{split}
& R_{Z_{ij}Z_{kj}}(2\zeta_i\frac{P}{4}c_{ik})^{\otimes ^{ikj}}\otimes \\
& R_{Z_{ij}}(-2\zeta_i\frac{P}{4}c_{ik})^{\otimes ^{ij}} \otimes  \\ 
& R_{Z_{kj}}(-2\zeta_i\frac{P}{4}c_{ik})^{\otimes ^{kj}}   
\end{split}
\end{equation}

This approach typically requires the overlap coefficients \(c_{ik}\) to be precomputed before constructing the ansatz circuit. However, instead of doing so, we aim to compute the coefficients as part of the quantum circuit itself.

To achieve this, we can take into account that the overlap coefficients act as a control mechanism for the rotations applied by the \(R_{zz}\) and  \(R_z\) gates. Specifically, if tasks $i$ and $k$ are in conflict, then $c_{ik} = 1$, and the corresponding penalty rotations are applied. Conversely, if there is no conflict ($c_{ik}=0$), no rotation is applied. 

This behavior allows us to replace the \(R_{zz}\) and  \(R_z\) gates with their controlled counterparts, \(\mathit{CR_{zz}}\) and  \(\mathit{CR_z}\) respectively. These gates are controlled by a set of qubits that represent the coefficients \(c_{ik}\), as shown in the following expression.

\begin{equation}
\begin{split}
&\mathit{CR}^{c_{ik}}_{Z_{ij}Z_{kj}}(2\zeta_i\frac{P}{4})^{\otimes ^{ikj}}\otimes \\
&\mathit{CR}^{c_{ik}}_{Z_{ij}}(-2\zeta_i\frac{P}{4})^{\otimes ^{ij}} \otimes \\
&\mathit{CR}^{c_{ik}}_{Z_{kj}}(-2\zeta_i\frac{P}{4})^{\otimes ^{kj}} 
\end{split}
\end{equation}

The circuit used to compute the control qubits is described in detail in the next section, where we present two different versions applicable to different scenarios.

\subsection{\(H_B\) ansatz} 

The \(H_B\) ansatz corresponding to the standard operator \(U (H_B, \beta_i) \) defined in equation~\ref{eq:uhb} consists on a single set of \(R_X\) gates applied to all the qubits representing the \(x_{ij}\) variables, that is:
\begin{equation}
    R_{X_{ij}}(2\beta_i)^{\otimes ^{ij}}
    \label{fhm}
\end{equation}

\section{Conflict Detection}
\label{sec:Conflict}

As we mentioned in the previous section, the overlapping coefficients \(c_{ik}\) will be included within the QTIS-QAOA ansatz as a set of ancillary qubits that controls the application of the constraint penalties encoded in the \(H_c\) Hamiltonian.   

In this work, we propose two different strategies for encoding the overlap information in the control qubits. The first approach—\textit{Full Quantum}—estimates task overlaps using a quantum circuit, which requires only minimal classical pre-processing to normalize the time intervals that define each task.

In contrast, the second approach—\textit{Classical}—relies on a classical algorithm to perform most of the computations before translating the results into the quantum circuit. Although this method incurs a higher classical computational cost, it typically requires fewer qubits and delivers greater accuracy than the \(Quantum\) method in most scenarios.

\subsection{Full Quantum variant for conflict detection}

The quantum variant of the circuit is defined on the basis of the following method to determine whether two tasks overlap in time.

Consider two tasks, \(T^i\) and \(T^k\), associated with the time intervals \( [t_s^i, t_e^i] \) and \( [t_s^k, t_e^k] \), where \( t_s^i, t_s^k \) denote the start times and \( t_e^i, t_e^k \) denote the end times. The tasks are said to overlap if the end of task $i$ is greater than the start of task $k$ and the end of task $k$ is greater than the start of task $i$, that is:

\begin{equation}
    t_e^i > t_s^k \wedge t_e^k > t_s^i
\end{equation}

Based on this condition, we can determine the overlap coefficients \( c_{ik} \) as follows:

\begin{equation}
c_{ik} = \begin{cases}
1 &  t_e^i - t_s^k > 0 \wedge  t_e^k - t_s^i > 0 \\
0 & \text{otherwise} \\
\end{cases}
\end{equation}

Using this definition, we construct a quantum circuit with one qubit \(q_{t}\) per task t, and an ancillary qubit \(qc_{ik}\) for each overlap coefficient \( c_{ik} \).  The state \(\left|qc_{ik}\right\rangle = \left|1\right\rangle \)  indicates that tasks \(i\) and \(k\) overlap, while the state \(\left|0\right\rangle \) represents the absence of overlap.

The qubits \(q_i\),\(q_k\) will be used to compute the probabilities of \(t_e^i - t_s^k > 0 \) and \(t_e^k - t_s^i > 0\). For that, we begin by applying a Hadamard gate to each \(q_i\) qubit to prepare them in the superposition state \(\left|+\right\rangle\). 

From this point, the differences \(t_e^i - t_s^k\), \(t_e^k - t_s^i\) are mapped as rotations around the Y-axis applied to qubits $i$ and $k$, respectively. 

\begin{equation}
\begin{split}
\left|i\right\rangle & =R_Y(2(t_e^i - t_s^k))\left|+\right\rangle \\ 
\left|k\right\rangle & =R_Y(2(t_e^k - t_s^i))\left|+\right\rangle
\end{split}
\label{ftransformation}
\end{equation}

\begin{figure}[ht]
    \centering
    \includegraphics[width=0.8\columnwidth]{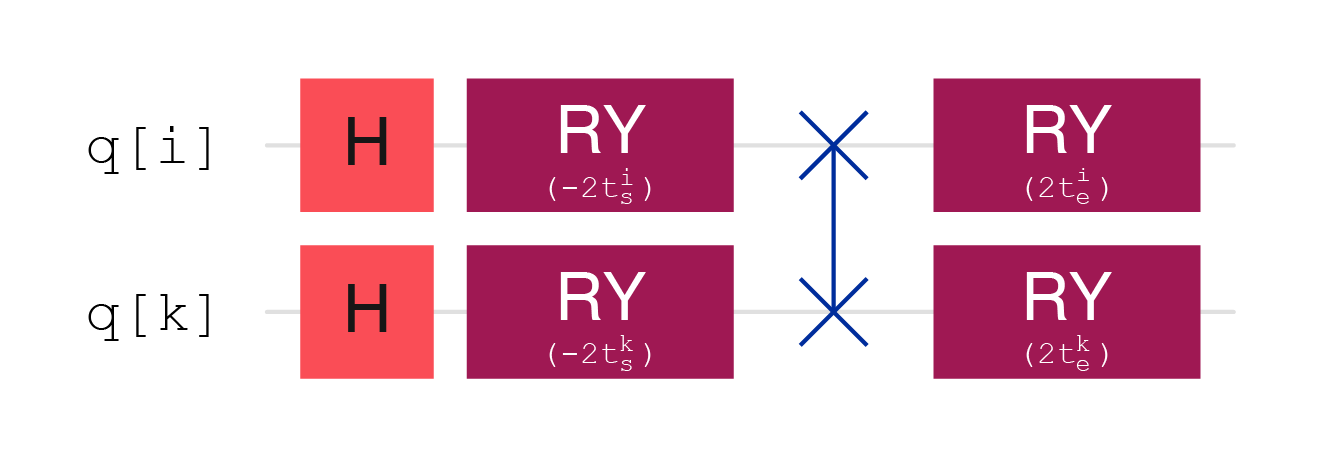}
    \caption{Circuit for time difference calculations}
    \label{fig:ts-te}
\end{figure}

The operation defined in the previous equation can be decomposed into a sequence of gates, starting with the application of the rotations \(R_Y(-2t_s^i)\) and \(R_Y(-2t_s^k)\)  to qubits \(i\), \(k\), respectively. This is followed by a SWAP gate between the two qubits, and finally by the application of \(R_Y(2t_e^i)\), \(R_Y(2t_e^k)\) rotation to the same qubits. This sequence is shown in Fig.~\ref{fig:ts-te}

\begin{figure*}[tb]
    \centering
    \includegraphics[width=\textwidth]{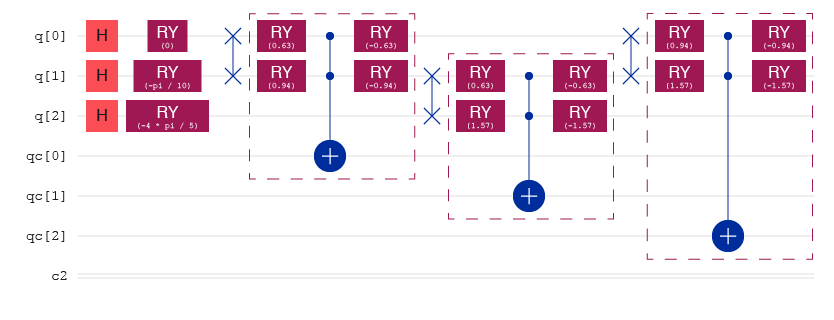}
    \caption{Quantum approach for the collision detection circuit with the damselfly gates between dashed lines boxes}
    \label{fig:colision_quantum}
\end{figure*}

\begin{align}
\left|ik\right\rangle &= 
\left[ R_Y(2 t_e^i) \otimes R_Y(2 t_e^k) \right] \cdot SWAP \cdot \nonumber \\
&\quad \left[ R_Y(-2t_s^i) \left|+\right\rangle_i \otimes R_Y(-2t_s^k) \left|+\right\rangle_k \right]
\label{eq:quantum_circuit}
\end{align}

\begin{figure}[ht]
    \centering
    \includegraphics[width=0.8\columnwidth]{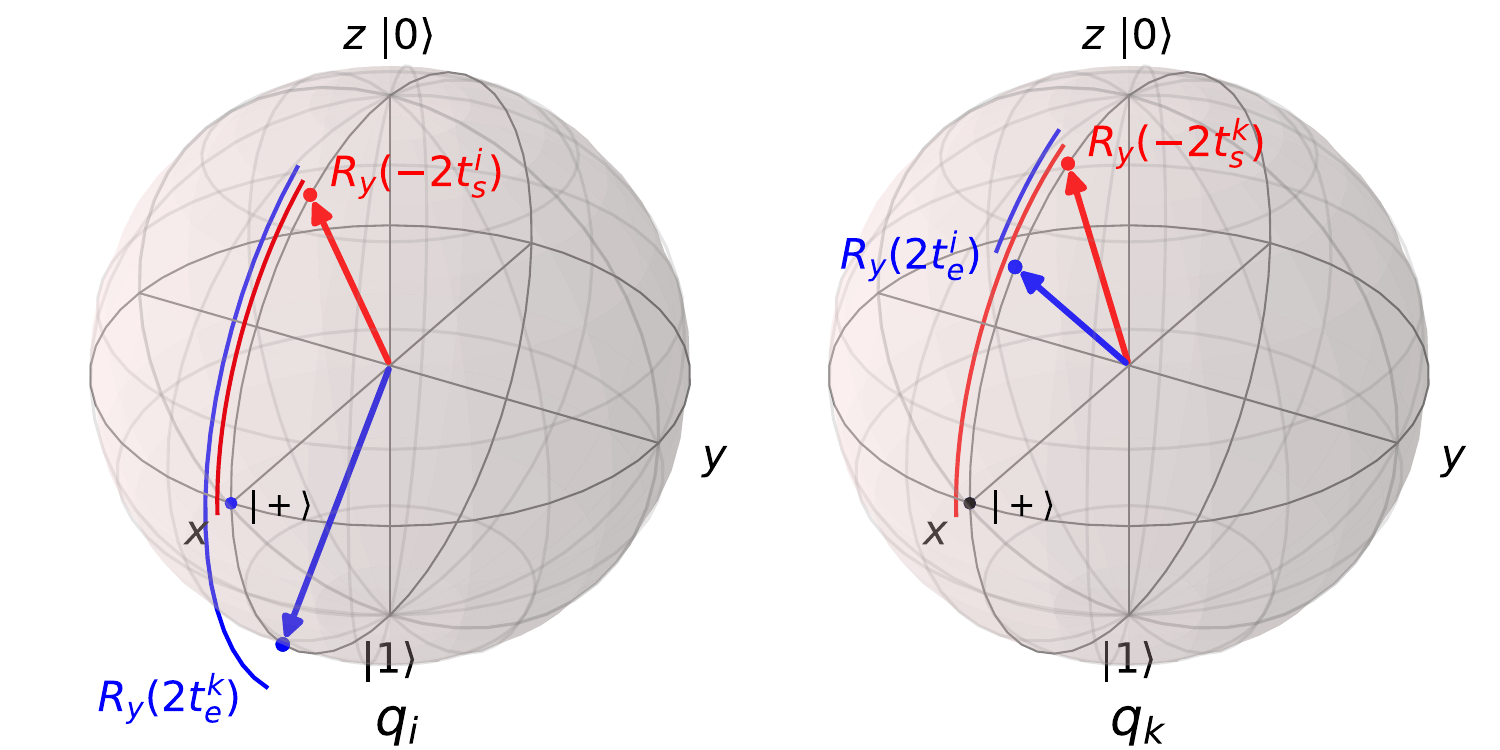}
    \caption{$R_Y$ rotations applied to qubits i, k from the superposition state $\left|+\right\rangle$}
    \label{fig:qubitsrotation}
\end{figure}

This computation requires that the interval times \(t_s^i, t_e^i \ \forall i\) be scaled in the range \([0, \pi/2]\). In Fig.~\ref{fig:qubitsrotation} we can see how the physical interpretation of these operations is directly related to the probabilities that \(q_i, q_k\) be in state \(\left|0\right\rangle\), \(t_e^i - t_s^k < 0\) and \(t_e^k - t_s^i < 0\), or in state \(\left|1\right\rangle\), \(t_e^i - t_s^k > 0\) and \(t_e^k - t_s^i > 0\).

Subsequently, the ancillary qubit \( qc_{ik} \), representing the overlap between i and k,  is computed using a Toffoli (CCNOT) gate controlled by \( q_i \) and \( q_k \), effectively implementing a logical AND between the two.

\begin{equation}
    \left|c_{ik}\right\rangle = \text{CCNOT}(\left|i\right\rangle\otimes\left|k\right\rangle )
\end{equation}


\subsection{Reduction of circuit complexity}

The number of qubits required to compute the overlap coefficients for the $N$ tasks can be reduced to:
\begin{equation}
   N_{qubits} = N + \frac{N(N-1)}{2} 
\end{equation}

This can be achieved by computing the coefficients in a sequential process. In each step, we calculate one of the coefficients \(c_{i,k}\) and, after applying the CCNOT gate, reverse the operation
\(
R_Y(2 t_e^i) \otimes R_Y(2 t_e^k)
\)
by applying \( R_Y \) rotations in the opposite direction to qubits \( i \) and \( k \). In the next step, we apply the SWAP gate again to the pair of qubits involved in the computation of the next coefficient \( c_{i' k'} \) and perform the corresponding rotations 
\(
R_Y(2 t_e^{i'}) \otimes R_Y(2 t_e^{k'})
\)
before applying the \(\mathit{CCNOT}(\left|i'\right\rangle\otimes\left|k'\right\rangle)\) gate. This iterative process continues until all coefficients have been calculated.

For practical purposes, we can define a custom gate for the transformation:
\begin{equation}
\begin{split}
    D = & [R_Y(-2 t_e^i) \otimes R_Y(-2 t_e^k)] \otimes \\
    &\mathit{CCNOT}(\left|i^l\right\rangle \otimes \left|k^l\right\rangle)\otimes \\
    &[R_Y(2 t_e^i) \otimes R_Y(2 t_e^k)]
\end{split}
\end{equation}

With this structure, which we call the \textit{Damselfly} gate, we can compute collision probabilities for all tasks by combining SWAP gates and Damselfly gates in a sequence of stages.

Figure~\ref{fig:colision_quantum} shows an example circuit for three tasks, in which the overlap coefficients are computed in three sequential gate stages, and where:

\begin{itemize}
    \item \( q[0], q[1], q[2] \) represent tasks 0, 1, and 2, respectively.
    \item \( qc[0] \) represents the overlap between tasks 0 and 1, \( c_{0,1} \).
    \item \( qc[1] \) represents \( c_{0,2} \).
    \item \( qc[2] \) represents \( c_{1,2} \).
\end{itemize}

\subsubsection{Limitations}

The presented approach can exhibit limitations in certain cases due to the probabilistic nature of quantum measurement outcomes.

The method is only suitable for interval sets where the differences \(t_e^i - t_s^k\) and \(t_e^k - t_s^i\) are sufficiently large to ensure that the coefficient \(c_{ik}\)  can be measured with a clear probability corresponding to either the \(\left|1\right\rangle\) or \(\left|0\right\rangle\) state. 

If this is not the case, the calculation of the initial rotations, as well as those performed within the \textit{damselfly} gates, would not be achievable through a simple scaling of the time intervals within the range \([0,2\pi]\). Instead, they would need to be specifically calculated for each step. Additionally, in such scenarios, the damselfly gates would lose their antisymmetric property, increasing the amount of classical pre-processing required and raising the classical computational cost.

\begin{figure*}[t!]
\centering
\newsavebox{\ClassicColission}
\begin{lrbox}{\ClassicColission}
\begin{minipage}{0.9\linewidth}
\small
\begin{verbatim}
def ClassicalConflictCircuit(task_intervals):

    differences = []
    for i in range(len(task_intervals)):
        interval1 = task_intervals[i]
        for k in range(i + 1, len(task_intervals)):
            interval2 = intervals[k]
            #t_s^k - t_e^i computation
            differences.append(interval2[0] - interval1[1])
    
    colisiones = np.copysign(1,differences)

    nQubits = len(colisiones)
    qRCollisions = QuantumRegister(nQubits, name="qC")
    cRCollisions = ClassicalRegister(nQubits, name="cC")
    qc = QuantumCircuit(qRCollisions, cRCollisions)

    # Quatum circuit generation
    for i in range(0, nQubits):
        qc.h(i)  # Qubit in supperposition  
        qc.ry(-colisiones[i]*np.pi/2, i) # Ry rotation

    return qc
\end{verbatim}
\end{minipage}
\end{lrbox}

\fbox{\usebox{\ClassicColission}}\\[1mm]

\caption{Qiskit code to generate classical conflict circuit variant}
\label{fig:qiskit}
\end{figure*}

\subsection{Classical variant for conflict detection}

The classical variant of the circuit for computing the overlap coefficients qubits is based on a different algorithm from the one used in the quantum variant. This approach is particularly suitable in scenarios where the quantum implementation cannot be applied due to the limitations discussed in the previous section.

In this case, the classical approach begins with a list of tasks that are pre-sorted by their start times. Given this ordered list, we can efficiently verify whether two tasks \( T^i = [t_s^i, t_e^i] \) and \( T^k = [t_s^k, t_e^k] \), where \( t_s^k \geq t_s^i \), overlap by checking the following condition:

\begin{equation}
    t_s^k - t_e^i < 0
\end{equation}

Using this method, we can define a quantum circuit that requires a total number of qubits equal to:

\begin{equation}
    n_{qubits} = \frac{N (N-1)}{2}, \quad N = \text{number of tasks}
\end{equation}

where each qubit \( qc_{ik} \) represents a pairwise interval comparison. 

The computation of each overlap coefficient \( c_{ik} \) begins by applying a Hadamard gate to place the qubit in superposition, followed by an \( R_Y \) rotation with an angle of \(-\frac{\pi}{2} \cdot \text{sign}(t_s^k - t_e^i)\). This rotation encodes whether the time difference between the start of task $k$ and the end of task $i$ is positive or negative, enabling the qubit to probabilistically reflect the overlap condition.

\begin{equation}
    R_Y\Big(-\frac{\pi}{2} \cdot \text{sign}(t_s^k - t_e^i)\Big)^{\otimes \frac{N (N-1)}{2}} \otimes H^{\otimes \frac{N (N-1)}{2}}
\end{equation}

\begin{figure}[ht]
    \centering
    \includegraphics[width=0.7\columnwidth]{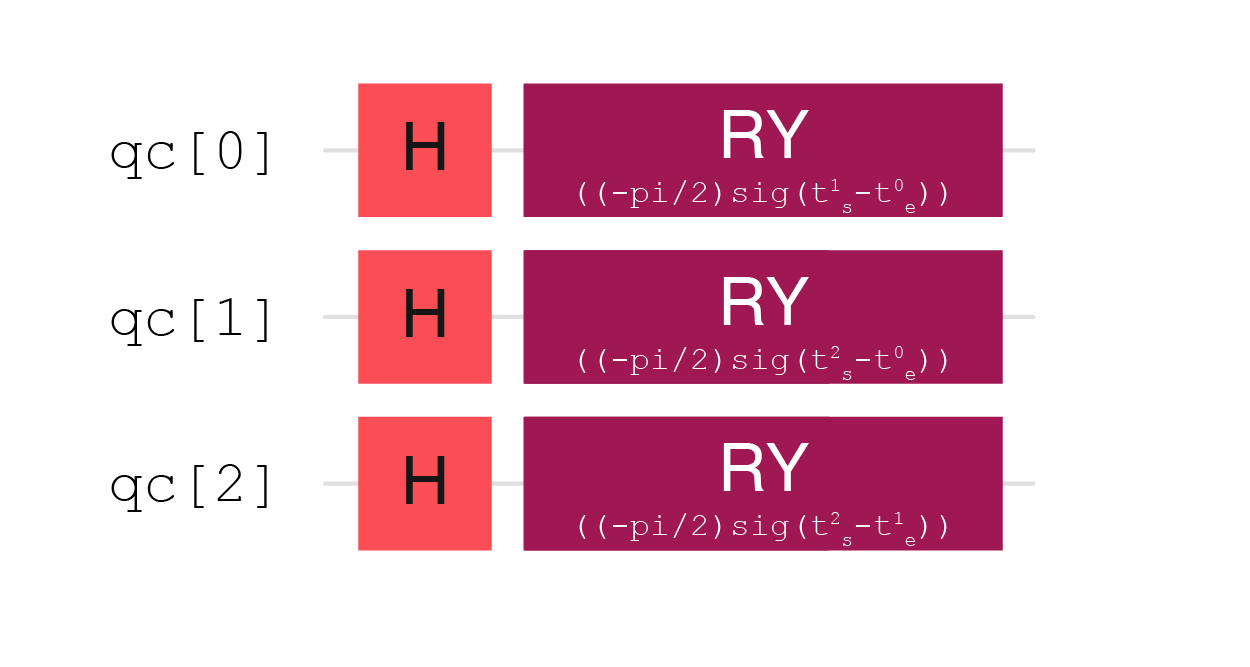}
    \caption{Classical approach for the collision detection circuit
    collision circuit for a set of three tasks.}
    \label{fig:classicalapproach}
\end{figure}

In Fig.~\ref{fig:qiskit} we can see the algorithm to build the classical variant of the conflict circuit implemented in \textit{Qiskit}. The result for an example with three tasks is shown in Fig.~\ref{fig:classicalapproach}.

\subsection{QTIS ansatz implementation}

With all elements already defined, we can now construct the complete QTIS ansatz by taking the Kronecker product of the QTIS-QAOA circuit and one of the collision circuits, along with the connections between the ancillary qubits $qc_{ik}$ and the controlled gates $R_z$ and $R_{zz}$ within the $H_c$ layers of the QTIS-QAOA ansatz. 

\begin{figure*}[ht]
    \centering
    \begin{subfigure}{\textwidth}
        \centering
        \includegraphics[width=\textwidth]{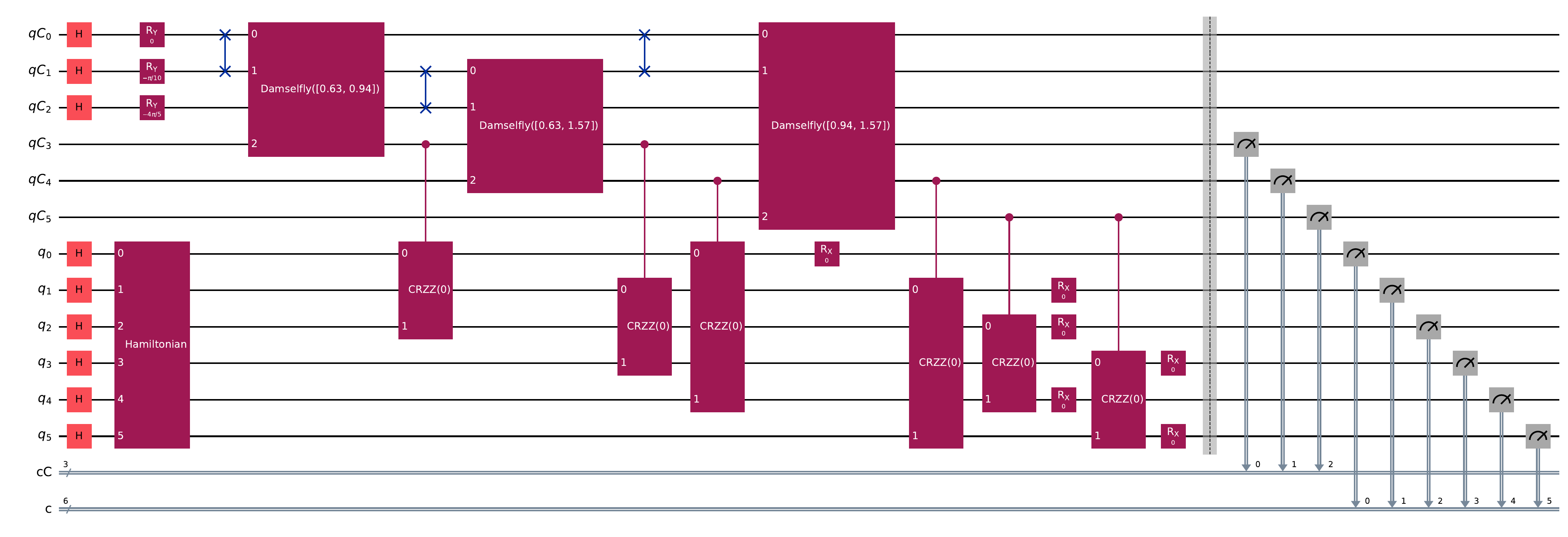}
        \caption{Quantum collision variant}
        \label{fig:qtiscircuitsub1}
    \end{subfigure}
    \begin{subfigure}{\textwidth}
        \centering
        \includegraphics[width=\textwidth]{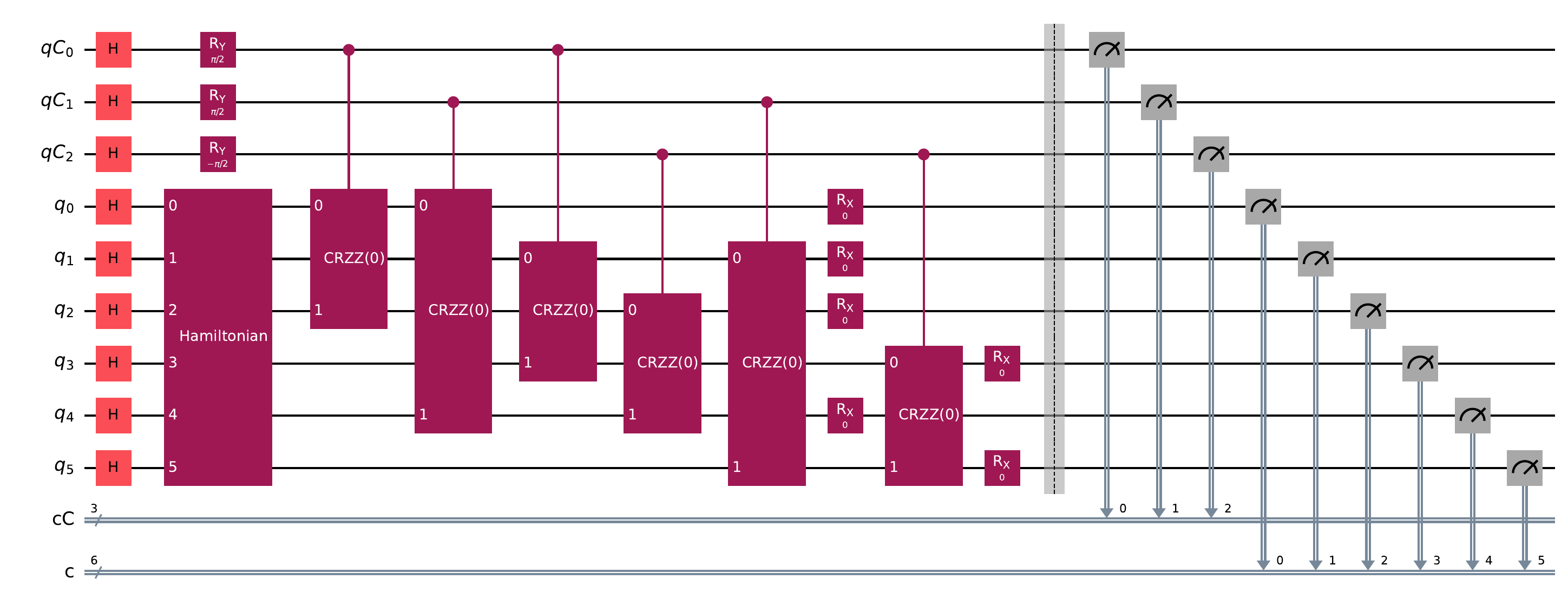}
        \caption{Classical collission variant}
        \label{fig:qtiscircuitsub2}
    \end{subfigure}
    \caption{QTIS implementation with Quantum (a) and Classical (b) variants for the conflict detection circuit.}
    \label{fig:qtiscircuit}
\end{figure*}

Fig.~\ref{fig:qtiscircuit} shows an example circuit for a three-task scheduling problem with a QAOA depth of
$L=1$, for both quantum and classical variants of the conflict circuit. In the diagrams, we can also observe that measurement gates have been added to both the $qc_{ik}$ qubits and to the qubits involved in the QTIS-QAOA circuit.

\section{Simulations and results}
\label{sec:Simulations}

To validate the QTIS algorithm, we conducted a series of tests with a dual objective: 

\begin{enumerate}
    \item Identify a suitable minimization strategy for the QTIS-QAOA ansatz that is capable of finding a valid task allocation among the available resources.
    \item Determine whether introducing a third parameter set, \( \vec{\zeta} \), specifically for the \(H_c\) circuit (instead of sharing the \(\vec{\gamma}\) parameters with the \(H_p\) circuit), leads to an improvement in the minimization process.
\end{enumerate}

\subsection{Test sets definition}

To perform the tests, we define six sets of three tasks, each with a different degree of overlap. The tasks in each set, as shown in Table~\ref{tab:intervals}, are scheduled using two available resources.

\begin{table}[ht]
    \centering
    \caption{Interval sets for testing}
    \begin{tabularx}{\columnwidth}{clX}
    \toprule
        SET & TASK INTERVALS & OVERLAPPING \\
    \midrule
        1 & [1,3],[1.5,4],[5,6] & task 1 and task 2 \\
        2 & [1,3],[1.5,8],[2,6] & all tasks are overlapped \\
        3 & [1,3],[1.5,4],[3.5,6] & task 1 and task 2, task 2 and task 3 \\
        4 & [1,2],[3,4],[5,6] & none tasks are overlapped \\
        5 & [1,2],[3,5],[4,6] & task 2 and task 3 \\
        6 & [1,5],[2,3],[4,6] & task 1 and task 2, task 1 and task 3 \\ 
    \bottomrule
    \end{tabularx}
    \label{tab:intervals}
\end{table}

For each set of tasks, we perform batches of 10 simulations using the three different minimization strategies defined in the next section: standard QAOA, T-QAOA\cite{ZhouLeo2020QAOA}, and HT-QAOA.  
For all strategies, we conduct two batches: one with \( \vec{\zeta} = \vec{\gamma}\) and another with \(\vec{\zeta} \neq \vec{\gamma}\), to evaluate which configuration best suits the problem defined by the test set.

In all cases, the QTIS-QAOA ansatz is simulated with a depth of \(L=10\) using the Qiskit Aer simulator and the Qasm backend. The number of shots per simulation is fixed at 100,000.

All these tests are initially performed using the classical variant for the conflict detection circuit. We will later compare the results obtained for the first interval set with those from the quantum variant of the conflict detection circuit.

At the end of each simulation, we measure the normalized energy of the solution:

\begin{equation}
    E_{norm} = \frac{E(\vec{\gamma}, \vec{\zeta}, \vec{\beta}, c_{ik}) - E_{min}}{E_{max} - E_{min}}
\end{equation}

where 

\begin{equation}
    E(\vec{\gamma}, \vec{\zeta}, \vec{\beta}, c_{ik}) = \left\langle\vec{\gamma}, \vec{\zeta}, \vec{\beta}, c_{ik}|H|\vec{\gamma}, \vec{\zeta}, \vec{\beta}, c_{ik} \right\rangle
\end{equation}

Using the normalized energy as a reference allows us to consistently compare the results across the different test sets.

\subsection{Minimization strategies}

As mentioned above, the QTIS algorithm is a variant of the QAOA algorithm and therefore requires a minimization process to obtain optimal values for the parameters \(\vec{\gamma}, \vec{\zeta_l}\), and \(\vec{\beta_l}\).
The minimization process, along with the initial parameter selection, greatly influences both the quality of the result and the computational complexity of the algorithm's execution.
In this work, we evaluate three different strategies: standard minimization, T-QAOA, and HT-QAOA.

\subsubsection{Standard minimization strategy}
The standard minimization strategy consists of evaluating the QTIS-QAOA circuit with the chosen depth of \(L = 10\) iteratively. In each iteration, the parameters \(\vec{\gamma}, \vec{\zeta_l}\) and \(\vec{\beta_l}\) are adjusted using the COBYLA algorithm to minimize the normalized energy of the expectation value:

\begin{equation}
\left\langle \vec{\gamma}, \vec{\zeta}, \vec{\beta} \right| H_P \left| \vec{\gamma}, \vec{\zeta}, \vec{\beta} \right\rangle
\end{equation}

where:

\begin{equation}
\begin{split}
    \left| \vec{\gamma}, \vec{\zeta}, \vec{\beta} \right\rangle = & U(H_B, \beta_L)U(H_c, \zeta_L)U(H_p, \gamma_L)\dots \\
    &U(H_B, \beta_1)U(H_c, \zeta_1)U(H_p, \gamma_1) \left|s\right\rangle 
\end{split}
\end{equation}

\begin{equation}
    \left|s\right\rangle = \left|+_1\right\rangle \dots \left|+_n\right\rangle
\end{equation}

The initial values of the parameters \(\vec{\gamma}, \vec{\zeta_l}\) and \(\vec{\beta_l}\) in this case are randomly assigned using a uniform distribution function in the interval \([0,\pi]\).

\subsubsection{T-QAOA strategy}

The T-QAOA strategy aims to obtain the parameters \(\vec{\gamma}, \vec{\zeta}\) and \(\vec{\beta}\) for a given circuit depth $L$, using as a starting point the parameters previously calculated for depth $L-1$.

In T-QAOA, we begin by applying a standard minimization, defined above, with a depth of \(L = 1\), and in successive stages, the standard minimization is reapplied while incrementally increasing the circuit depth \(L\), so that at each stage \(L = l\), the parameters \(\vec{\gamma}_l, \vec{\zeta}_l\), and \(\vec{\beta}_l\) are initialized as follows:

\begin{equation}
\begin{split}
    \vec{\gamma}_{1..l-1}^l &= \vec{\gamma}_{1..l-1}^{l-1}\\
    \vec{\gamma}_l^l &= \vec{\gamma}_{l-1}^{l-1}    
\end{split}
\end{equation}

\begin{equation}
\begin{split}
    \vec{\zeta}_{1..l-1}^l &= \vec{\zeta}_{1..l-1}^{l-1}\\
    \vec{\zeta}_l^l &= \vec{\zeta}_{l-1}^{l-1}    
\end{split}
\end{equation}

\begin{equation}
\begin{split}
    \vec{\beta}_{1..l-1}^l &= \vec{\beta}_{1..l-1}^{l-1}\\
    \vec{\beta}_l^l &= 0    
\end{split}
\end{equation}

\subsubsection{HT-QAOA strategy}

The third minimization strategy, which we introduce in this work, is referred to as HT-QAOA.
This strategy is influenced by the principles of quantum annealing and homotopy optimization described in \cite{KunduAkash2024Hhqa}.

According to these principles, we can reach a target Hamiltonian starting from an auxiliary Hamiltonian by gradually varying a parameter \(\alpha\) as follows:
\begin{equation}
H = \alpha \cdot H_{obj} + (1-\alpha) \cdot H_m 
\end{equation}

As \(\alpha\) approaches 1, the influence of the auxiliary Hamiltonian diminishes while the contribution of the target Hamiltonian increases.

QAOA draws inspiration from this principles by alternating between the problem Hamiltonian \(H_P\) and the mixing Hamiltonian \(H_B\), each parameterized by
\(\vec{\gamma}\) and \(\vec{\beta}\), respectively.
Through successive optimization stages, and as the circuit depth increases, it is commonly observed that the parameters \(\vec{\beta}\) tends to approach zero, while \(\vec{\gamma}\) converges toward \(\pi\).

Building on this idea, the proposed HT-QAOA strategy consists of two steps:

\begin{enumerate}
    \item Perform an initial standard minimization with circuit depth \(L = 1\).
    \item Perform a second standard minimization with the full circuit depth L so that the parameters \(\vec{\gamma}\), \(\vec{\zeta}\), and \(\vec{\beta}\) are initialized using a linear interpolation from the parameters obtained in the first step. Specifically, \(\gamma_l\) and \(\zeta_l\) are initialized within the ranges \([\gamma_1,\pi]\) and \([\zeta_1,\pi]\), respectively, while \(\beta_l\) is initialized within the range \([\beta_1, 0]\).    
\end{enumerate}  

\subsection{Results}

As previously described, in order to compare the three strategies, we conducted batches of ten simulations per algorithm for each set of tasks, considering both cases: \(\vec{\gamma} \neq \vec{\zeta}\) and \(\vec{\gamma} = \vec{\zeta}\). 

The simulations were carried out in a virtualized environment with the specifications indicated in table~\ref{tab:envspecs}. 

\begin{table}[ht]
    \centering
    \caption{Simulation Environment Specifications}
    \begin{tabular}{lp{4cm}}
    \toprule
    Parameter & Specification \\
    \midrule
    CPU & 8 vCPUs — Intel(R) Xeon(R) Silver 4314 @2.40GHz \\
    RAM & 32 GB \\
    Operating System & Ubuntu Linux (64-bit) \\
    Virtualization Platform & VMWare ESXi 8.0 \\
    \bottomrule
    \end{tabular}
    \label{tab:envspecs}
\end{table}

The result of each simulation is the normalized energy, calculated using the corresponding $E_{min}$ and $E_{max}$ values for each task set. Both, $E_{min}$ and $E_{max}$ are defined by the Hamiltonian of the problem and its associated QUBO formulation, as described in equation \eqref{eq:fnqubo}. In this study,  $E_{min}$ is the same in all sets and has a value of -3\footnote{This is the value of the QUBO in equation \eqref{eq:fnqubo} for three taks and two resources when no penalties are applied}. On the other hand, $E_{max}$ is estimated by considering the QUBO in equation\eqref{eq:fnqubo} under the assumption that all binary variables \(x_{ij}\) \eqref{eq:fnqubo} are set to 1. The values of $E_{min}$ and $E_{max}$ for all sets are shown in table~\ref{tab:energybounds}.

\begin{table}[ht]
    \centering
    \caption{Minimum and Maximum Energy Values for Each Task Set}
    \begin{tabular}{ccccccc}
        \toprule
         & Set 1 & Set 2 & Set 3 & Set 4 & Set 5 & Set 6 \\
        \midrule
        $E_{min}$ & -3 & -3 & -3 & -3 & -3 & -3 \\
        $E_{max}$ & 29 & 57 & 43 & 15 & 29 & 43 \\
        \bottomrule
    \end{tabular}
    \label{tab:energybounds}
\end{table}

\subsubsection{Comparison between minimization strategies}

\begin{table}[ht]
    \centering
    \caption{Comparison between minimization strategies \\ (Normalized Energy for \(\vec{\gamma} \neq \vec{\zeta}\) and \(\vec{\gamma} = \vec{\zeta}\))}
    \begin{tabular}{c c c c}
    \toprule
        Set & Standard & T-QAOA & HT-QAOA \\
        \midrule
        \multicolumn{4}{c}{\(\vec{\gamma} \neq \vec{\zeta}\)} \\
        \midrule
        1 & 0.04 & 0.07 & 0.10 \\
        2 & 0.19 & 0.15 & 0.17 \\
        3 & 0.16 & 0.12 & 0.14 \\
        4 & 0.00 & 0.00 & 0.00 \\
        5 & 0.04 & 0.08 & 0.07 \\
        6 & 0.19 & 0.12 & 0.14 \\
        \midrule
        MEAN & 0.10 & 0.09 & 0.10 \\
        \midrule
        \multicolumn{4}{c}{\(\vec{\gamma} = \vec{\zeta}\)} \\
        \midrule
        1 & 0.10 & 0.10 & 0.11 \\
        2 & 0.22 & 0.16 & 0.20 \\
        3 & 0.21 & 0.14 & 0.15 \\
        4 & 0.00 & 0.00 & 0.00 \\
        5 & 0.13 & 0.10 & 0.11 \\
        6 & 0.21 & 0.13 & 0.18 \\
        \midrule
        MEAN & 0.145 & 0.105 & 0.125 \\
        \bottomrule
    \end{tabular}
    \label{tab:comparissonstrategies}
\end{table}

Table \ref{tab:comparissonstrategies} presents the mean normalized energy obtained for each simulation batch, as well as the overall mean energy achieved by each minimization strategy across all task sets.

\begin{figure*}[htbp]
  \centering
  \begin{minipage}[b]{0.49\textwidth}
    \includegraphics[width=\textwidth]{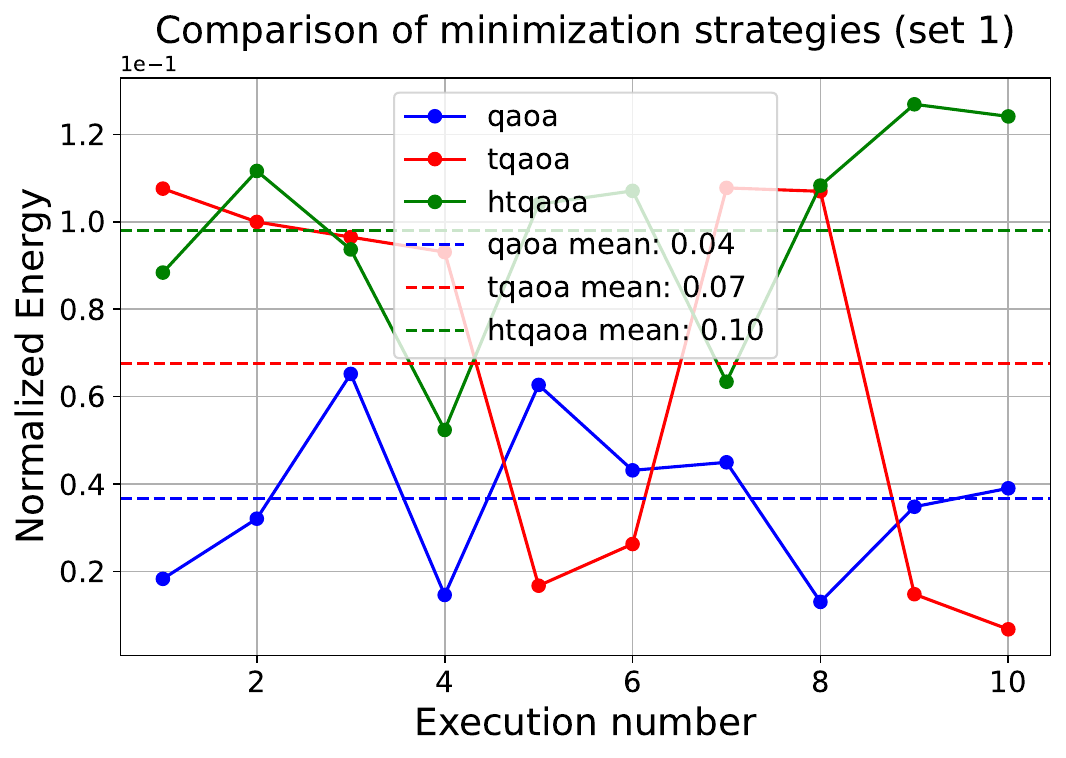}
  \end{minipage}
  \hfill
  \begin{minipage}[b]{0.49\textwidth}
    \includegraphics[width=\textwidth]{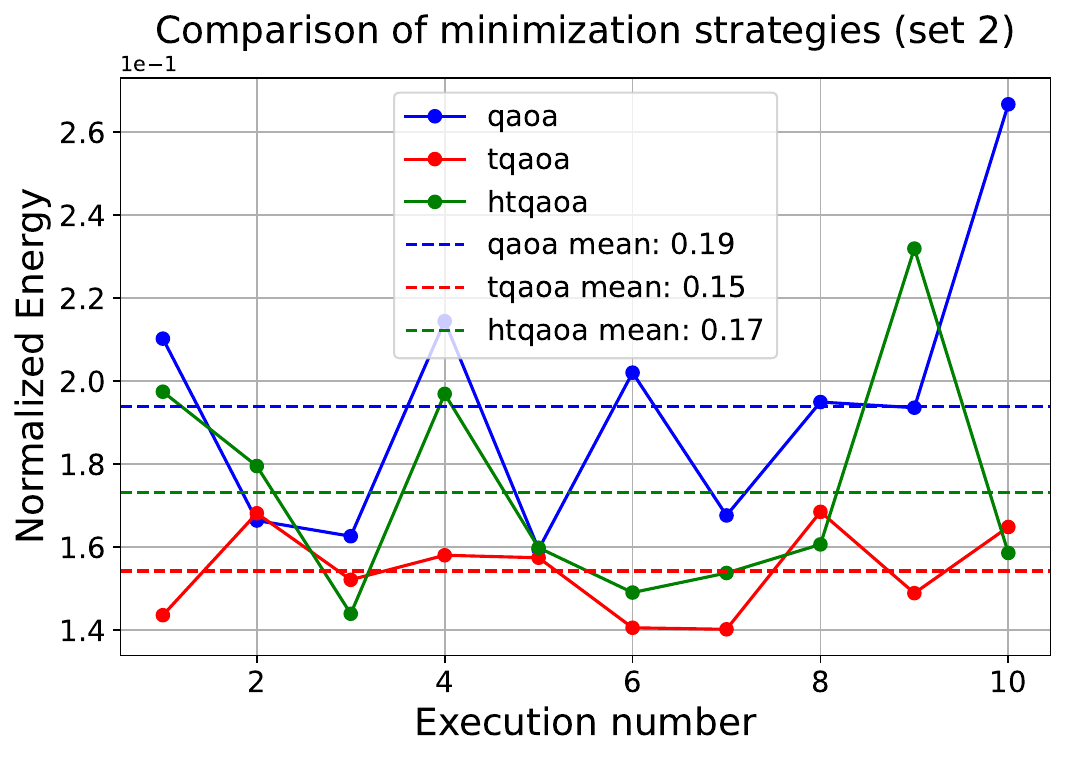}
  \end{minipage}

  \vspace{2ex}

  \begin{minipage}[b]{0.49\textwidth}
    \includegraphics[width=\textwidth]{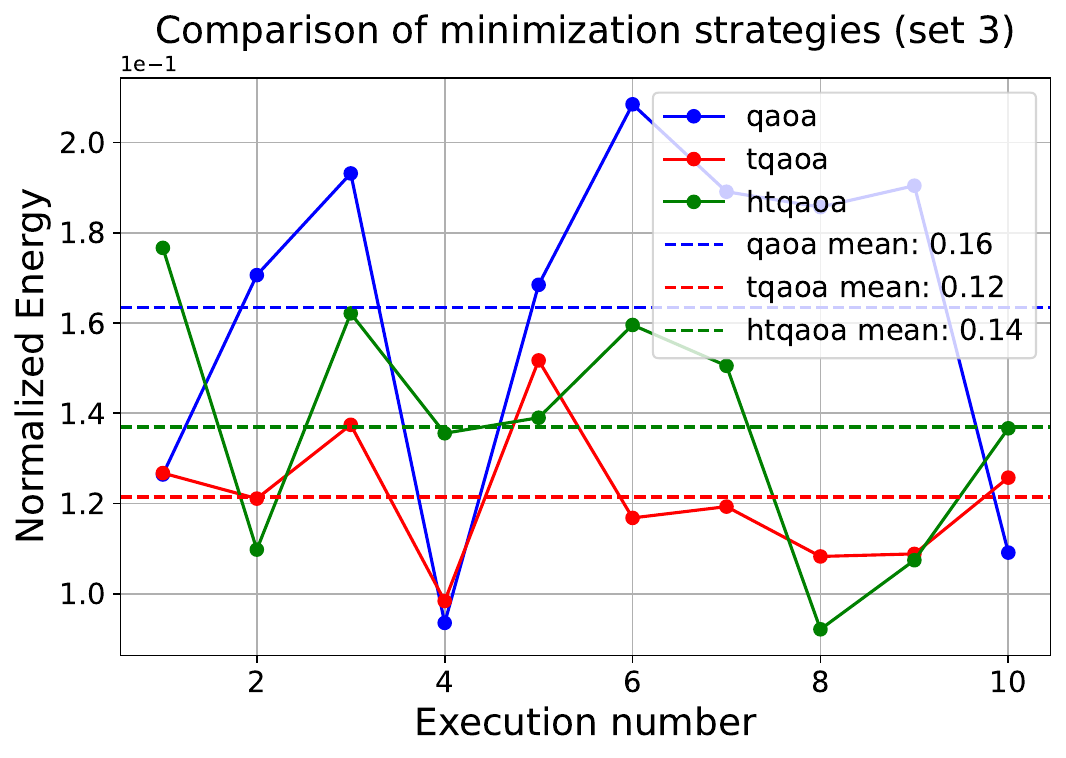}
  \end{minipage}
  \hfill
  \begin{minipage}[b]{0.49\textwidth}
    \includegraphics[width=\textwidth]{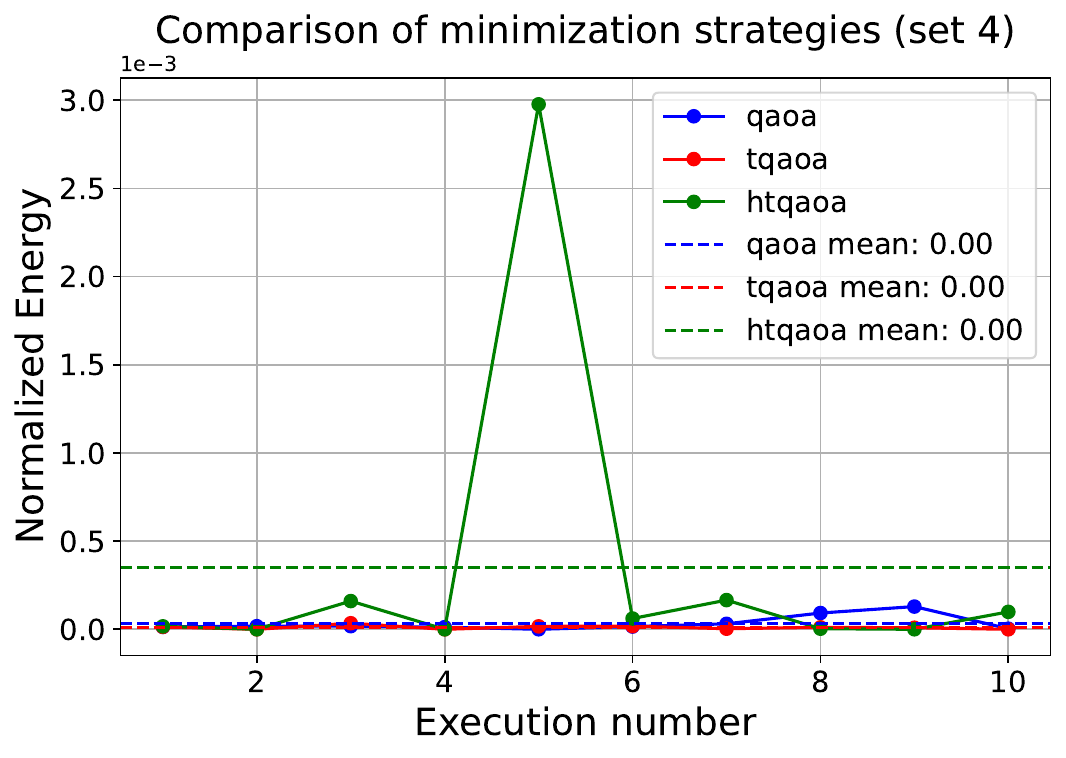}
  \end{minipage}

  \vspace{2ex}

  \begin{minipage}[b]{0.49\linewidth}
    \includegraphics[width=\linewidth]{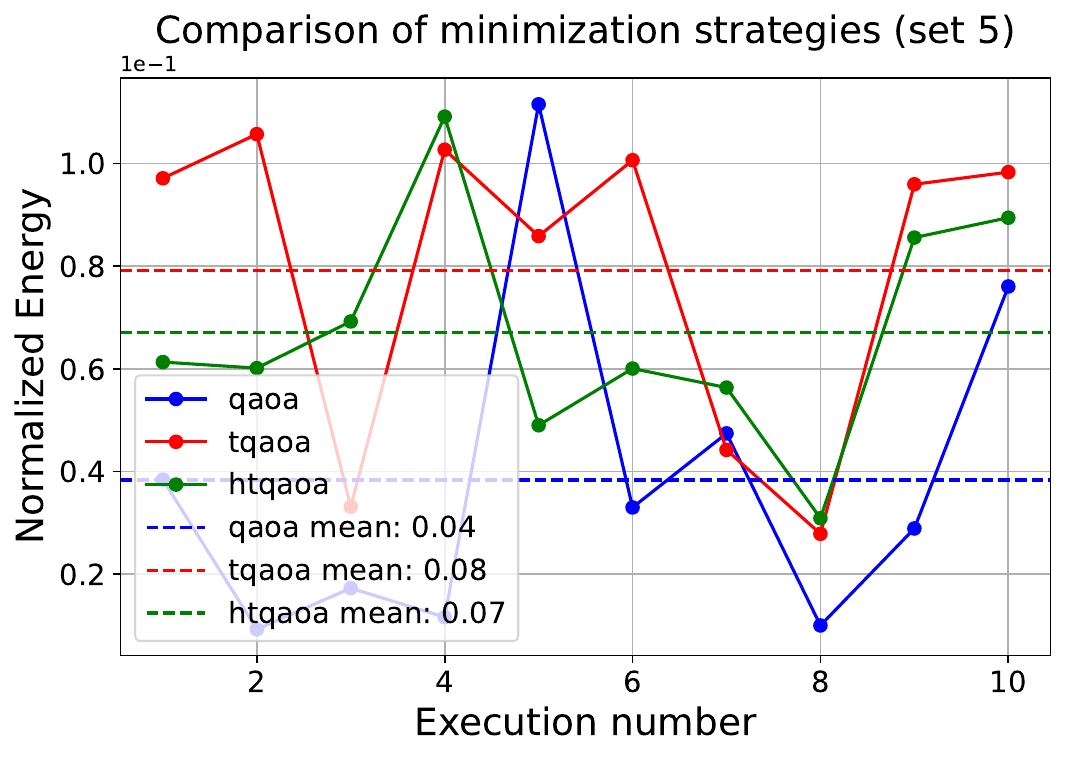}
  \end{minipage}
  \hfill
  \begin{minipage}[b]{0.49\linewidth}
    \includegraphics[width=\linewidth]{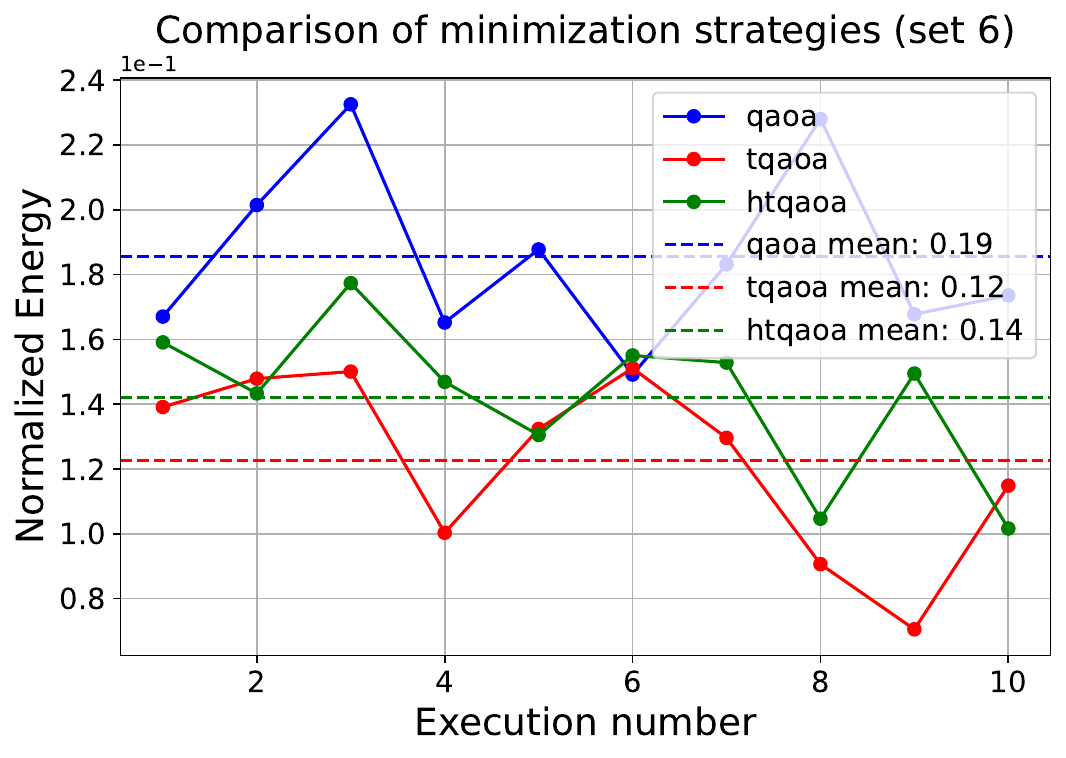}
  \end{minipage}

  \caption{Results for the three minimization strategies on the six interval sets.}
  \label{fig:strategies_comparison}
\end{figure*}

According to the mean values for the normalized energy of each strategy, T-QAOA is the strategy that obtains the best result with the lowest mean energy. Followed by HT-QAOA and the standard minimization. This can be seen in Fig.~\ref{fig:strategies_comparison}, where a graph is shown that compares the execution batches of the three strategies for each set with \( \gamma \neq \zeta \). 

\begin{figure}[ht]
    \centering
    \includegraphics[width=\columnwidth]{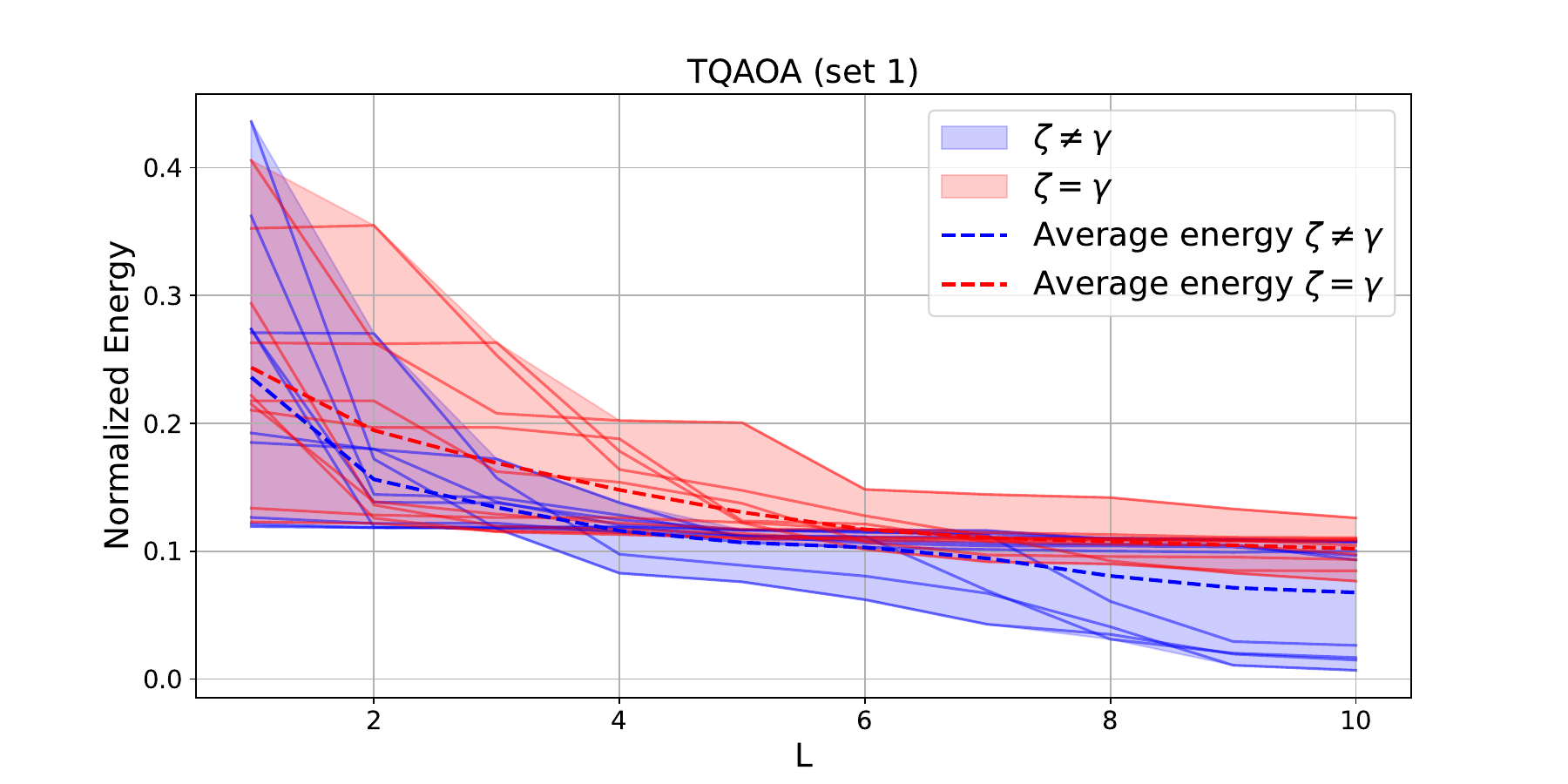}
    \caption{Normalized energy obtained with the T-QAOA strategy using 10 steps, for \(\vec{\gamma} \neq \vec{\zeta}\) and \(\vec{\gamma} = \vec{\zeta}\), in set 1}
    \label{fig:comparisontqaoagammazeta}
\end{figure}

In all cases, the strategy HT-QAOA sits between the standard and T-QAOA, even when T-QAOA is worse than the standard strategy. The mean execution times of each are indicated in table \ref{tab:execution_time}, where we can see that the execution time of HT-QAOA is similar to the standard strategy, while T-QAOA is between 3 and 4 times slower than the other two in these examples.  

\begin{table}[ht]
    \centering
    \caption{Mean execution time for the different minimization strategies (in seconds)}
    \begin{tabular}{cccc}
    \toprule
        Set & Standard & T-QAOA & HT-QAOA \\
        \midrule
        \multicolumn{4}{c}{\(\vec{\gamma} \neq \vec{\zeta}\)} \\
        \midrule
        1 & 52.42 & 177.97 & 49.10 \\
        2 & 42.12 & 167.73 & 43.31 \\
        3 & 46.71 & 169.70 & 44.49 \\
        4 & 44.00 & 157.60 & 49.53 \\
        5 & 59.08 & 187.47 & 49.74 \\
        6 & 46.27 & 168.44 & 47.58 \\
        \midrule
        Mean & 48.44 & 171.48 & 47.29 \\
        \midrule
        \multicolumn{4}{c}{\(\vec{\gamma} = \vec{\zeta}\)} \\
        \midrule
        1 & 35.88 & 115.53 & 34.62 \\
        2 & 25.94 & 111.93 & 28.50 \\
        3 & 29.36 & 117.18 & 30.36 \\
        4 & 27.66 & 106.84 & 32.40 \\
        5 & 30.73 & 112.48 & 32.47 \\
        6 & 27.66 & 112.14 & 31.50 \\
        \midrule
        Mean & 29.54 & 112.68 & 31.64 \\
        \bottomrule
    \end{tabular}
    \label{tab:execution_time}
\end{table}

With these data on execution time, to determine whether it is worth using the T-QAOA strategy in a more complex scenario, we could apply the standard strategy followed by HT-QAOA and if the energy of HT-QAOA is smaller we can try T-QAOA if we need a better result.

\subsubsection{Comparison between \(\gamma \neq \zeta\) and \(\gamma = \zeta\) }

The mean normalized energy measured along the different simulations with the three minimization strategies (indicated in table \ref{tab:comparissonstrategies}) leads us to think that using three different parameters set \(\vec{\gamma}, \vec{\zeta} , \vec{\beta}\) to characterized the Hamiltonians \( H_p, H_c\) and \(H_m\) with \(\vec{\gamma} \neq \vec{\zeta}\) is a better option than using only two different parameters \((\vec{\gamma} = \vec{\zeta})\) if we want to obtain the best results in the minimization process. 

In Fig.~\ref{fig:comparisontqaoagammazeta} we can see in detail the difference between using two or three parameters in the T-QAOA strategy applied to the task set 1. In each step, the mean normalized energy of the ten simulations that we have carried out with the set is larger if we use two parameters than if we use three. 

Finally, in Appendix 2, we present box plots for each task set that compare the normalized energy measurements obtained in the different simulations for each strategy, both for the case where \(\vec{\gamma} \neq \vec{\zeta}\) and where 
\((\vec{\gamma} = \vec{\zeta})\).

These diagrams clearly show that the median value of the normalized energy is consistently lower when \(\vec{\gamma} \neq \vec{\zeta}\), across all strategies and task sets.

\section{Conclusions}
\label{sec:Conclusion}

This article has addressed the task scheduling problem, which belongs to the class of NP-hard combinatorial optimization problems. 
In most cases, these problems are modeled as a subclass of the Job Shop Scheduling Problem (JSSP), whose objective is to execute all tasks in the shortest possible time using limited resources. 
However, this work has a slightly different objective: tasks are subject to strict temporal constraints and must be executed at specific times. Therefore, the goal is not focused on minimizing the total execution time but rather on fitting all the tasks within the limited set of available resources.

To tackle this challenge, the problem has been formulated as a QUBO model.
A novel variant of the QAOA algorithm has been proposed, which has been named \mbox{QTIS-QAOA}.
In order to construct the {\it ansatz}, three operators have been incorporated: \(U(H_p, \gamma_i)\), \(U(H_c, \zeta_i)\), and \(U(H_B, \beta_i)\), in contrast to the original QAOA, which relies on only two operators, \(U(H_P, \gamma_i)\) and \(U(H_B, \beta_i)\). This extension is intended to allow the operator \(U(H_c, \zeta_i)\) to take into account elements computed in a new section within the quantum circuit itself.

The Hamiltonians \(H_p\) and \(H_c\) can be considered a decomposition of the general \(H_P\) Hamiltonian obtained from the QUBO model of the problem.  

The operator \(U(H_c, \zeta_i)\) encapsulates the section of the QUBO that includes penalties for assigning two tasks to the same resource at the same time. These penalties depend on collision coefficients, \(c_{ik}\). In our approach, these \(c_{ik}\) coefficients are computed using a set of ancillary qubits, with two proposed quantum circuits for different scenarios. 
Subsequently, these qubits act as controls for the application of penalty terms in the \(U(H_c, \zeta_i)\) operator.

Computing the collision coefficient using a fully quantum approach may result in a loss of accuracy in certain cases. This is why a classical preprocessing alternative for determining the coefficients is also proposed in case these limitations are present.

To evaluate the effectiveness of the proposals, multiple simulations have been
carried out on six task sets, using three different minimization strategies:
Standard QAOA and T-QAOA, and a novel one, HT-QAOA, which constitutes an additional contribution of this work. 
HT-QAOA initializes parameters \(\vec{\gamma},\vec{\zeta},\vec{\beta}\) by
extrapolating from the results of an initial simulation with circuit depth
\(L=1\), following principles from quantum annealing and homotopy optimization.

Experimental results show that HT-QAOA achieves intermediate performance between standard QAOA and T-QAOA. Nevertheless, it maintains an execution time comparable to that of the standard version. This behavior suggests the potential to use low-depth for determining the most suitable optimization strategy for a given instance before scaling up.
The results also indicate that using three distinct parameter sets—\(\vec{\gamma},\vec{\zeta},\vec{\beta}\)—with \(\vec{\gamma} \neq \vec{\zeta}\) leads to better quality solutions compared to using \(\vec{\gamma} = \vec{\zeta}\), although at the cost of increased computational effort.
Future work will investigate scalability to larger task graphs and alternative Hamiltonian decompositions to enhance performance and generalization.

\appendix
\section{Ising}
\label{app_ising}

In this appendix we are going to calculate the Ising model for the \(H_c\) Hamiltonian. 

As we have seen, \(H_c\) are derived from the constraints \(x_{ij}x_{kj}c_{ik} = 0 \quad \forall i,j,k>i\), which establish that tasks cannot be executed simultaneously on the same resource.

The QUBO penalties corresponding to these constraints are in the formula:  

\begin{equation}
    \mathit{QUBO_{H_c}} =  \sum_j\sum_k\sum_iP(x_{ij}x_{kj}c_{ik})^2
\label{eq:fnqubohc_appendix}
\end{equation}

To calculate the Ising model, we have to apply the QUBO to Ising transformation: \(x_{ij} = \frac{(1-s_{ij})}{2}, x_{kj} = \frac{(1-s_{kj})}{2}\)

If we focus on the quadratic term, we can see that:

\begin{equation}
\begin{split}
    (x_{ij}x_{kj}c_{ik})^2 = & (\frac{1-s_{ij}}{2}\frac{1-s_{kj}}{2})^2c_{ik}^2 \\
    = & \frac{1}{4^2}(1-s_{ij}-s_{kj}+s_{ij}s_{kj})^2c_{ik} \\
    = &\frac{1}{4^2}(1-s_{ij}-s_{kj} + s_{ij}s_{kj} \\
    &-s_{ij}+s_{ij}s_{ij} +s_{ij}*s_{kj} - s_{ij}s_{ij}s_{kj}\\
    &-s_{kj}+s_{kj}s_{ij}+s_{kj}s_{kj}-s_{ij}s_{kj}s_{kj}\\
    &+s_{ij}s_{kj}-s_{ij}s_{ij}s_{kj}-s_{kj}s_{ij}s_{kj}\\
    &+s_{ij}s_{kj}s_{ij}s_{kj})c_{ik} \\
    = &\frac{1}{4^2}(1-s_{ij}-s_{kj} + s_{ij}s_{kj} \\
    &-s_{ij} + 1 + s_{ij}s_{kj} - s_{kj} \\
    &-s_{kj} + s_{kj}s_{ij}  + 1 -s_{ij} \\
    &+s_{ij}s_{kj}-s_{kj} -s_{ij}+ 1)c_{ik}\\
    = &\frac{1}{4^2}(4-4s_{ij}-4s_{kj}+4s_{ij}s_{kj})c_{ik}\\
    = &\frac{1}{4}(1-s_{ij}-s_{kj}+s_{ij}s_{kj})c_{ik}\\
\end{split}    
\label{fhcising}    
\end{equation}

So, if we substitute the previous results into \(\mathit{QUBO_{H_c}}\) expression, we can obtain the final Ising model for the \(H_c\) Hamiltonian as: 

\begin{equation}
    H_c = \sum_j\sum_k\sum_i\frac{P}{4}(1-s_{ij}-s_{kj}+s_{ij}s_{kj})c_{ik}
\end{equation}

\section{Box diagrams}
\label{app_box_diagrams}

\begin{figure*}[htbp]
  \centering
  \begin{minipage}[b]{0.49\textwidth}
    \includegraphics[width=\textwidth]{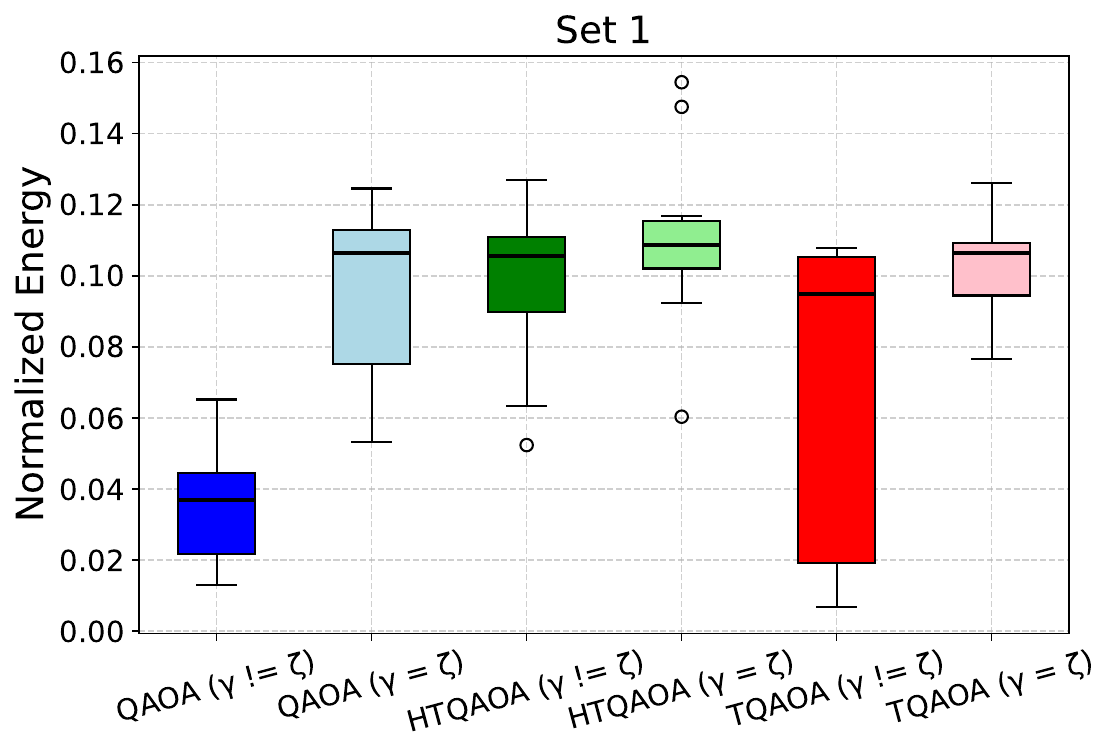}
  \end{minipage}
  \hfill
  \begin{minipage}[b]{0.49\textwidth}
    \includegraphics[width=\textwidth]{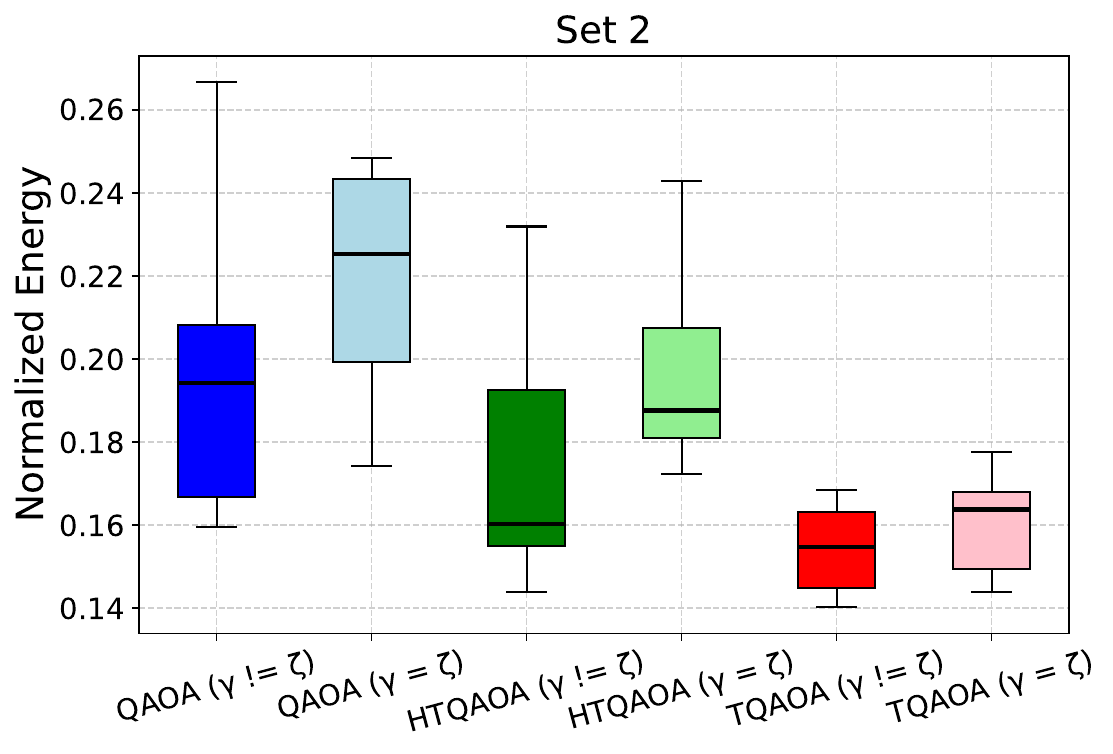}
  \end{minipage}

  \vspace{2ex}

  \begin{minipage}[b]{0.49\textwidth}
    \includegraphics[width=\textwidth]{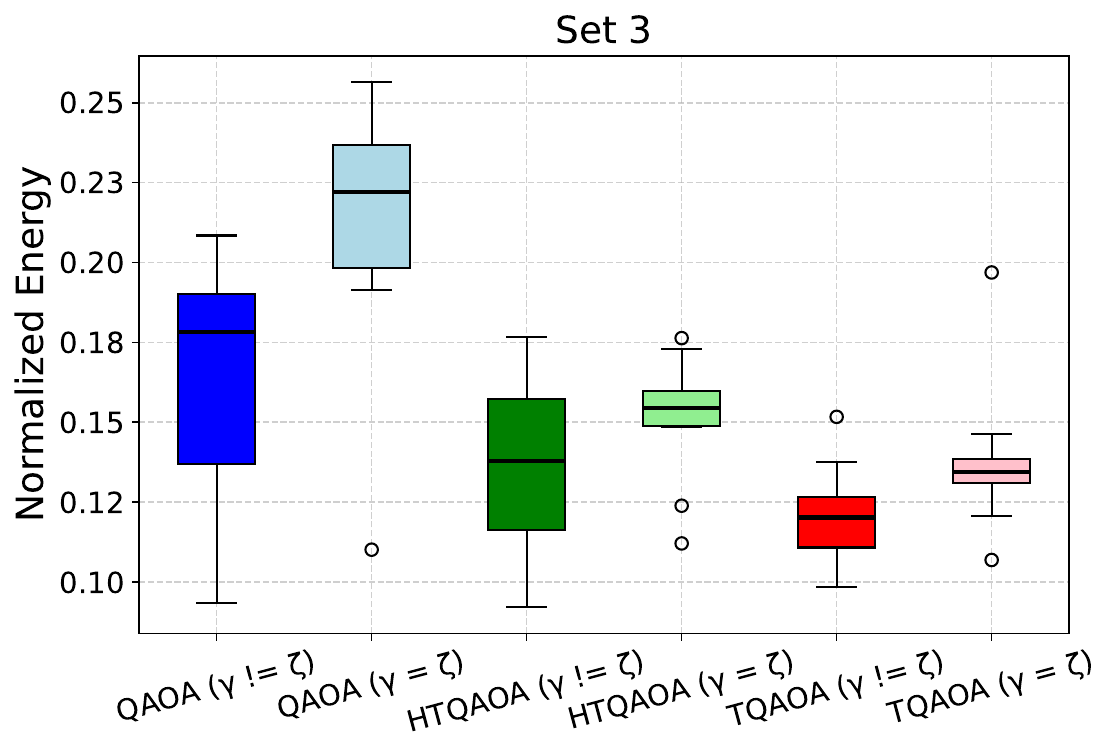}
  \end{minipage}
  \hfill
  \begin{minipage}[b]{0.49\textwidth}
    \includegraphics[width=\textwidth]{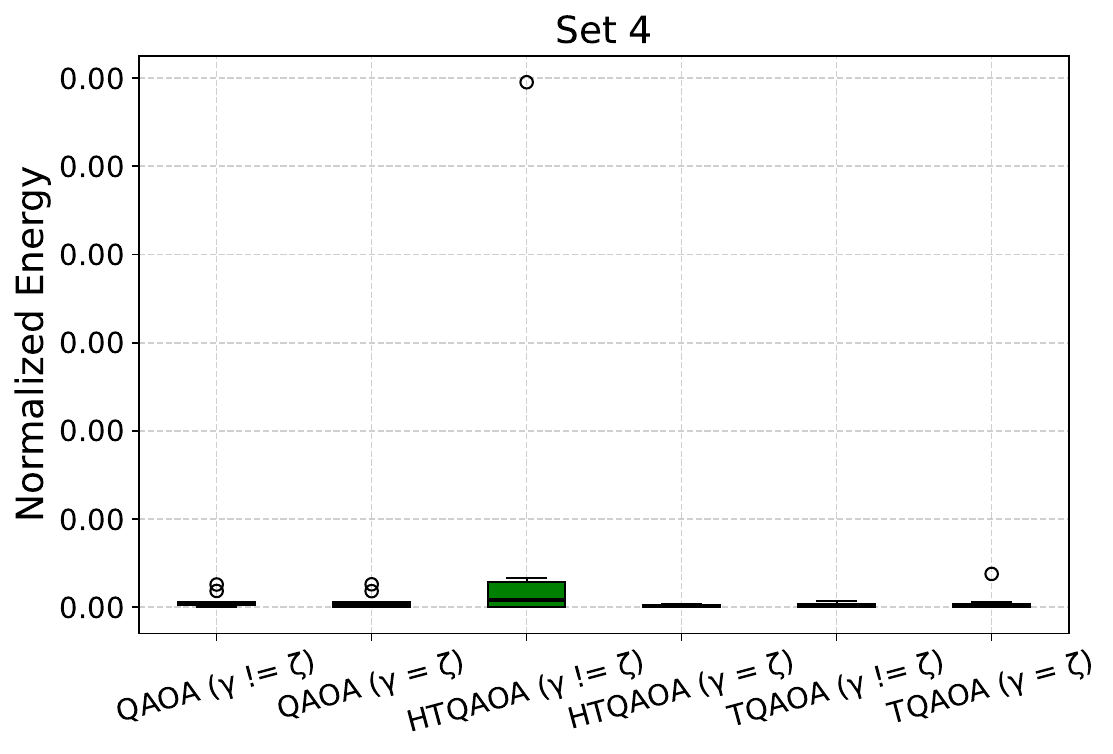}
  \end{minipage}

  \vspace{2ex}

  \begin{minipage}[b]{0.49\linewidth}
    \includegraphics[width=\linewidth]{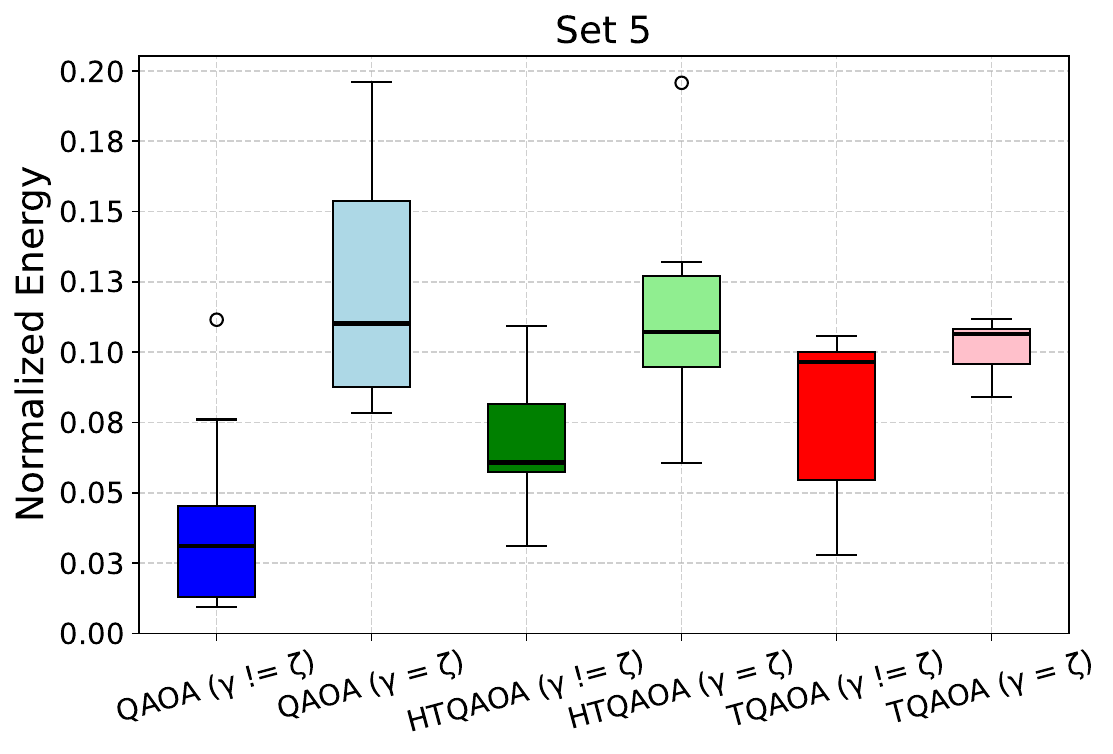}
  \end{minipage}
  \hfill
  \begin{minipage}[b]{0.49\linewidth}
    \includegraphics[width=\linewidth]{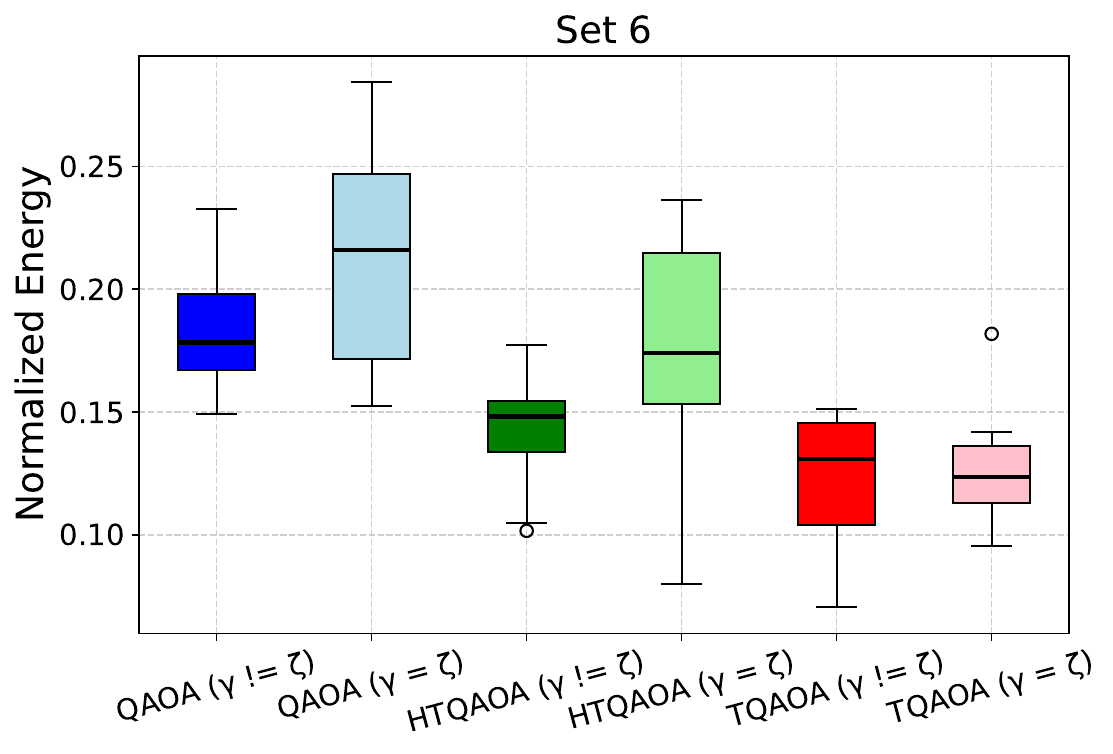}
  \end{minipage}

  \caption{Boxplot for Task Sets 1-6}
  \label{fig:boxdiagrams}
\end{figure*}

The box plots showing the distribution of normalized energy values obtained over 10 simulations for each task set and each minimization strategy--under both configurations \(\vec{\gamma} \neq \vec{\zeta}\) and 
\((\vec{\gamma} = \vec{\zeta})\)--are presented in Figure~\ref{fig:boxdiagrams}. 

\end{document}